\documentclass[superscriptaddress, twocolumn, amsmath, amssymb, aps,pra, letterdraft, notitlepage,longbibliography]{revtex4-1}
\usepackage{graphicx,graphics,epsfig,subfigure,times,bm,bbm,amssymb,amsmath,amsfonts,amsthm,mathrsfs,MnSymbol}
\usepackage[matrix,frame,arrow]{xypic}
\usepackage[normalem]{ulem}
\usepackage{slashed}
\usepackage{color}
\usepackage[usenames,dvipsnames,svgnames,table]{xcolor}
\usepackage[english]{babel}
\usepackage{verbatim}
\definecolor{darkblue}{rgb}{0.0,0.0,0.3}
\usepackage[colorlinks=true,
            linkcolor=red,
            urlcolor= darkblue,
            citecolor=blue]{hyperref}

\newcommand{\bea}{\begin{eqnarray}}
\newcommand{\eea}{\end{eqnarray}}

\begin{document}
\title{Dissipation-engineering of nonreciprocal quantum dot circuits: An input-output approach} 

\author{Junjie Liu}
\address{Department of Chemistry and Centre for Quantum Information and Quantum Control,
University of Toronto, 80 Saint George St., Toronto, Ontario M5S 3H6, Canada}
\author{Dvira Segal}
\address{Department of Chemistry and Centre for Quantum Information and Quantum Control,
University of Toronto, 80 Saint George St., Toronto, Ontario M5S 3H6, Canada}
\address{Department of Physics, 60 Saint George St., University of Toronto, Toronto, Ontario M5S 1A7, Canada}

\begin{abstract}
Nonreciprocal effects in nanoelectronic devices offer unique possibilities for manipulating 
electron transport and engineering quantum electronic circuits for information processing purposes. 
However, a lack of rigorous theoretical tools is hindering this development.  
Here, we provide a general input-output description
of nonreciprocal transport in solid-state quantum dot architectures, based on 
quantum optomechanical analogs. 
In particular, we break reciprocity between coherently-coupled quantum dots 
by dissipation-engineering in which these (so-called) primary dots are mutually coupled to  auxiliary, damped quantum dots. 
We illustrate the general framework in two representative  multiterminal noninteracting
models, which can be used as building blocks for larger circuits. 
Importantly, the identified optimal conditions for nonreciprocal behavior hold even in the presence of 
additional dissipative effects that result from local electron-phonon couplings. 
Besides the analysis of the scattering matrix, we show that a nonreciprocal coupling induces unidirectional electron flow in the resonant transport regime. 
Altogether, our analysis provides the formalism and working principles towards the realization of 
nonreciprocal nanoelectronic devices. 
\end{abstract}

\date{\today}

\maketitle

\section{Introduction}

The ability to break reciprocity in nanoscale devices is crucial for quantum information processing and telecommunication, 
as the nonreciprocal functionality can protect active elements against extraneous noise. 
Reciprocity may be broken in various ways, such as by introducing asymmetry and nonlinearity, as in electrical and thermal diodes.  
Recently, significant attention has been given to engineering nonreciprocal transmission amplitudes
in quantum optomechanical systems based on the interference between coherent couplings and dissipative effects
\cite{Hafezi.12.OP,Ranzani.14.NJP,Ranzani.15.NJP,Metelmann.15.PRX,Ruesink.16.NC,Xu.16.PRA,Xu.17.PRA,Xu.17.PRAa,Shen.16.NP,Bernier.17.NC,Peterson.17.PRX,Fang.17.NP,Barzanjeh.18.PRL,Malz.18.PRL,Xu.19.N}. 
%
The analogous effect for electron transport is also highly desirable, 
as proposals for quantum information processing and computing platforms are based on electronic networks 
such as quantum dot arrays \cite{Loss.98.PRA} or superconducting qubit circuits \cite{Devoret.13.S}. 
However, engineering nonreciprocal behavior in electronic systems---based 
on balancing coherent and dissipative couplings---remains almost unexplored, besides a recent study on a specific double quantum dot setup \cite{Malz.18.PRB}. One concrete challenge is that for electron flow, 
 one needs to control the charge 
current, which integrates over electrons in the full bias window, rather than only focus on the transmission probability at a particular energy. Much remains therefore to be learned about how to engineer nonreciprocal interactions in general electronic settings. 
Particularly, it is desirable to establish a simple yet general theoretical tool fitting for this task. 

Here, we employ a recent theoretical advance, 
termed the generalized input-output method (GIOM) \cite{Liu.20.A,Liu.20.B} to address this challenge, and describe nonreciprocal 
nanoelectronic devices.
Within the framework of the GIOM, the influence of fermionic environments exerted on electrons is encoded in 
generalized input and output fields, and the system dynamics is described by Heisenberg-Langevin equations (HLEs), 
thereby establishing a complete analogy to theoretical descriptions of quantum optomechanical systems. 
Building upon this input-output picture, we 
generate nonreciprocal couplings in electronic systems through dissipation engineering \cite{Metelmann.15.PRX}.
%
Focusing on solid-state quantum dot architectures, 
a simple yet general strategy to induce nonreciprocal coupling between quantum dots
lies in building interference between coherent (Hamiltonian) inter-dot couplings and dissipative interactions.
This interference is controlled by an applied, tunable magnetic field,
which builds complex-valued coherent tunneling elements \cite{Malz.18.PRB}.

We employ this mechanism in our setup by introducing primary and auxiliary quantum dots.
The primary quantum dots construct the circuit of interest between two electrodes, termed `left' ($L$) and `right' ($R$).
To control the directionality of electrons towards say the $R$ electrode, the primary dots
are mutually coupled to auxiliary quantum dots, which experience strong dissipation (loss) induced by additional
fermionic reservoirs. 
In the large damping limit, one can adiabatically eliminate the auxiliary dots,
resulting in an effective dissipative interaction
between primary quantum dots. 
Nonreciprocal coupling is generated
by balancing this induced dissipative interaction against the inherent coherent hopping processes, 
and the directionally of electron transmission at the $L$ and $R$ interfaces
is controlled by an external magnetic field. 
Figs. \ref{fig:fig2} and \ref{fig:fig4} present three-dot and four-dot examples with two primary dots ($d_1$, $d_2$) 
and one or two auxiliary dots ($a_1$, $a_2$).

To illustrate the utility and generality of our  approach,
we study two minimal models consisting of three or four dots, which mimic two different
situations in which {\it direct} coherent hopping between the two primary dots may or may not be present, 
as in Figs \ref{fig:fig2} and \ref{fig:fig4}, respectively. 
Larger networks can be built based on these minimal building blocks. 
In the noninteracting scenario, we easily identify optimal conditions for efficient nonreciprocal charge transmission from the 
{\it exact} HLEs and the scattering matrix obtained from the input-output relation. 
This analysis expands the scope of a previous study \cite{Malz.18.PRB},
and provides more insights into the underlying mechanism in general settings. 
As HLEs allow a simple yet exact solution for the charge current, we examine nonequilibrium configurations 
and demonstrate a unidirectional charge transport due to nonreciprocal couplings,
without engineering the spectral densities of the fermonic reservoirs \cite{Mascarenhas.19.PRB,Damanet.19.PRL}. 
We further address the impact of electron-phonon couplings on nonreciprocal interaction engineering:
By introducing a polaron-dressed scattering matrix, we show that the 
conditions for optimal nonreciprocal charge transmission are unaffected by the presence of local electron-phonon coupling.

Altogether, in this study we (i) describe a general framework for building non-reciprocity in electronic circuits,
analyzed through the scattering matrix and the charge current, 
(ii) study two central models with nonreciprocal behavior
induced by countering direct or indirect coherent coupling by an engineered dissipation, and
(iii) rigorously show that local elecrton-phonon interaction does not affect conditions for non-reciprocity.

The paper is organized as follows. 
In Sec. \ref{s:1}, we introduce the general setup and the input-output scheme based on the GIOM. 
In Sec. \ref{s:2}, we illustrate how to generate nonreciprocal behavior
through dissipation-engineering using two minimal models. 
Signatures of nonreciprocal interactions in nonequilibrium charge transport are 
demonstrated in Sec. \ref{s:3}. 
We address the impact of electron-phonon coupling and non-engineered dissipation on
nonreciprocity in Sec. \ref{s:4} and summarize in Sec. \ref{s:5}.

\section{Model and input-output method for charge transport}\label{s:1}

The general setup includes coherently-coupled quantum dots and fermionic environments,
described by the Hamiltonian (hereafter, $\hbar=1$, $e=1$, $k_B=1$ and Fermi energy $\epsilon_F=0$)
\begin{equation}\label{eq:HH}
H~=~H_{coh}+H_{diss}.
\end{equation}
Here, $H_{coh}$ is the coherent system Hamiltonian, and it includes 
both primary and auxiliary quantum dots. It includes fermionic operators and only quadratic terms.
We specify this term in Sec. \ref{s:2}. 
We are interested in controlling the charge current entering and leaving the primary dots.
Auxiliary dots serve to prepare the nonreicprocal interaction.
For simplicity, for each dot we only include a single electronic level
with annihilation operators $\{d_n\}$ and $\{a_m\}$ for the primary and auxiliary dots, respectively.
$H_{diss}$ includes dissipative-damping terms of all dots, in the form of particle loss to 
independent metallic leads,
\bea
H_{diss}~&=&
\sum_{k,n}\Big[\epsilon_{kn}^{d}c_{kn}^{d,\dagger}c_{kn}^{d}+t_{kn}^{d}(c_{kn}^{d,\dagger}d_n
+d_n^{\dagger}c_{kn}^{d})\Big]
\nonumber\\
&+&
\sum_{k,m}\Big[\epsilon_{km}^{a}c_{km}^{a,\dagger}c_{km}^{a}+t_{km}^{a}(c_{km}^{a,\dagger}a_m
+a_m^{\dagger}c_{km}^{a})\Big]
\eea
The first row describes the coupling of the primary dots (counted by $n$) to the source and drain metal leads.
$c_{kn}^{d}$ annihilates an electron with energy $\epsilon_{kn}^{d}$ in the lead that couples to the $n$ primary dot
with $t_{kn}^{d}$ the tunneling rate.  The superscript `d' highlights that these dissipation terms affect primary dots.
The second row describes the coupling of auxiliary dots (counted by $m$) to their own metal leads, which induce dissipation. 
The definitions for auxiliary dots are similar to those of primary dots, with the  superscript `a' marking terms related to the auxiliary dots.
The damping effects induced by the different sets of metallic leads are characterized 
by the spectral densities
$\Gamma_n(\epsilon)=\pi\sum_k(t_{kn}^{d})^2\delta(\epsilon-\epsilon_{kn}^{d})$ 
and $\kappa_m(\epsilon)=\pi\sum_k(t_{km}^{a})^2\delta(\epsilon-\epsilon_{km}^{a})$.
Note that in our work, we follow conventions from the charge transport literature, 
with $\Gamma$ as a width parameter induced by the source and drain electrodes, and $\kappa$ serving as a
damping rate constant due to auxiliary reservoirs. 
Notably, in the quantum optics literature the opposite convention is used, with $\kappa$ as the coupling rate to the input and output
physical ports, and $\Gamma$ the engineered-bath damping rate \cite{Metelmann.15.PRX}. 

We now briefly introduce the input-output equations for the above electronic system. 
More details on the GIOM can be found in Ref. \cite{Liu.20.A, Liu.20.B}. 
Interpreting as initial conditions, generalized input fields from metallic leads are defined as
\bea
d_{n,in}(t) &=& \frac{1}{\sqrt{2\pi}}\sum_kt_{kn}^de^{-i\epsilon_{kn}^d(t-t_0)}c_{kn}^d(t_0),
\nonumber\\
a_{m,in}(t) &=& \frac{1}{\sqrt{2\pi}}\sum_kt_{km}^ae^{-i\epsilon_{km}^a(t-t_0)}c_{km}^a(t_0),
\label{eq:3}
\eea
with $t_0$ the initial time at which the dynamical evolution begins. 
Generalized output fields, $d_{n,out}$, $a_{m,out}$,
relate to input fields through the following input-output relations
\bea\label{eq:in_out}
d_{n,out}(t) &=& d_{n,in}(t)-i\sqrt{\frac{2}{\pi}}\Gamma_n d_{n}(t),\nonumber\\
a_{m,out}(t) &=& a_{m,in}(t)-i\sqrt{\frac{2}{\pi}}\kappa_m a_m(t).
\eea
We emphasize that in the wide band limit \cite{Wingreen.89.PRB},
the dot-lead hybridization energies,
 $\Gamma_n$ and $\kappa_m$ are treated in an exact manner in the GIOM. 
For simplicity, we set $\Gamma_n=\Gamma$ and $\kappa_m=\kappa$ hereafter. 
For our purposes, we consider the strong  damping limit for the auxiliary dots, such that $\kappa\gg\Gamma$. 

Defining column vectors $\boldsymbol{F}_{\beta}=(\{d_{n,\beta}\},\{a_{m,\beta}\})^T$ ($\beta =in,~out$), 
$\boldsymbol{O}=(\{d_n\},\{a_m\})^T$ and a diagonal matrix $\boldsymbol{K}=\mathrm{diag}(\{\Gamma\},\{\kappa\})$, 
the above boundary conditions can be recasted into
\begin{equation}\label{eq:io_relation}
\boldsymbol{F}_{out}(t)~=~\boldsymbol{F}_{in}(t)-i\sqrt{\frac{2}{\pi}}\boldsymbol{K}\cdot\boldsymbol{O}(t)
\end{equation}
Within the framework of the GIOM, dynamical evolution of an arbitrary system operator,
$\mathcal{O}$, is described by a so-called Heisenberg-Langevin equation (HLE),
\begin{equation}
\label{eq:eom_o}
\dot{\mathcal{O}}~=~ i[H_{coh},\mathcal{O}]_{-}-i\sum_{n}\mathbb{L}_{\pm}^n-i\sum_m\mathbb{A}_{\pm}^m,
\end{equation}
where 
\bea
\mathbb{L}_{\pm}^n&\equiv&\mp\left(i\Gamma d_{n}^{\dagger}+\sqrt{2\pi}d_{n,in}^{\dagger}\right)[\mathcal{O},d_{n}]_{\pm}
\nonumber\\
&+&[\mathcal{O},d_{n}^{\dagger}]_{\pm}\left(-i\Gamma d_{n}+\sqrt{2\pi}d_{n,in}\right),\nonumber\\
\mathbb{A}_{\pm}^m &\equiv&\mp\left(i\kappa a_m^{\dagger}+\sqrt{2\pi}a_{m,in}^{\dagger}\right)[\mathcal{O},a_m]_{\pm}
\nonumber\\
&+&[\mathcal{O},a_m^{\dagger}]_{\pm}\left(-i\kappa a_m+\sqrt{2\pi}a_{m,in}\right).
\eea
The top signs apply if $\mathcal{O}$ is a fermionic operator; the bottom signs apply if 
$\mathcal{O}$ is bosonic. We have defined $[A,B]_-\equiv[A,B]$ and $[A,B]_+\equiv\{A,B\}$ representing quantum commutator and anti-commutator, respectively. For the vector $\boldsymbol{O}$, we thus have \cite{Liu.20.A}
\begin{equation}\label{eq:eom_general}
\dot{\boldsymbol{O}}(t)=\boldsymbol{M}\cdot \boldsymbol{O}(t)+\boldsymbol{C}\cdot\boldsymbol{F}_{in}(t), 
\end{equation}
where we have introduced a time-independent drift matrix $\boldsymbol{M}$  
and a coefficient matrix $\boldsymbol{C}$ whose detailed forms are model-dependent. 

To engineer nonreciprocal interactions between two primary dots, 
one should find a {\it directionality condition} in which one dot is influenced by the other, but not vice versa, 
in the spirit of cascaded quantum systems \cite{Carmichael.93.PRL,Gardiner.93.PRL,Metelmann.15.PRX}. 
Moreover, for nonreciprocal interactions to be efficient in quantum signal transmission, 
we should further tune the system in such a way that the reflection of injected electrons at primary dots can be suppressed. 
This latter condition is referred to as 
{\it impedance matching condition}. 

We point out that the directionality condition can be obtained by adiabatically eliminating auxiliary degrees of 
freedom involved in HLEs for primary dots in the large damping limit. 
However, the impedance matching condition can not be inferred from HLEs. 
To obtain both conditions simultaneously, one should resort to the scattering matrix 
$\tilde{\boldsymbol{S}}[\omega]$ in the Fourier space, which relates input and output fields 
through $\tilde{\boldsymbol{F}}_{out}[\omega]=\tilde{\boldsymbol{S}}[\omega]\cdot\tilde{\boldsymbol{F}}_{in}[\omega]$,
 as was done in quantum optomechanical systems 
\cite{Hafezi.12.OP,Ranzani.14.NJP,Ranzani.15.NJP,Metelmann.15.PRX,Ruesink.16.NC,Xu.16.PRA,Shen.16.NP,Bernier.17.NC,Peterson.17.PRX,Fang.17.NP,Barzanjeh.18.PRL,Malz.18.PRL,Xu.19.N}.
Our GIOM indeed offers such a scattering matrix description. 
To this end, we consider Eqs. (\ref{eq:io_relation}) and (\ref{eq:eom_general}) 
in the frequency domain via Fourier transform, $\tilde{\boldsymbol{F}}_{out}[\omega]=\tilde{\boldsymbol{F}}_{in}[\omega]-i\sqrt{\frac{2}{\pi}}\boldsymbol{K}\cdot\tilde{\boldsymbol{O}}[\omega]$ and $\tilde{\boldsymbol{O}}[\omega]=(-i\omega\boldsymbol{I}-\boldsymbol{M})^{-1}\boldsymbol{C}\tilde{\boldsymbol{F}}_{in}[\omega]$ 
with $\boldsymbol{I}$ the identity matrix. 
Together, we obtain the scattering matrix
\begin{equation}
\label{eq:smatrix}
\tilde{\boldsymbol{S}}[\omega]~=~\boldsymbol{I}-i\sqrt{\frac{2}{\pi}}\boldsymbol{K}\cdot(-i\omega\boldsymbol{I}-\boldsymbol{M})^{-1}\cdot\boldsymbol{C},
\end{equation}
without actually solving the coupled dynamical evolution problem, Eq. (\ref{eq:eom_general}). 
An efficient nonreciprocal interaction between  primary dots $j$ and $k$ corresponds 
to maximizing the forward transmission coefficient 
$|\tilde{S}_{jk}|^2$ while minimizing the reverse transmission $|\tilde{S}_{kj}|^2$,
 as well as suppressing the reflections $|\tilde{S}_{jj}|^2$ and $|\tilde{S}_{kk}|^2$ with 
impedance matching. 
Hence, Eq. (\ref{eq:smatrix}) provides a general recipe for engineering nonreciprocal interactions in quantum dot systems.
This treatment is exact (under the wide band limit): 
Unlike quantum master equations of motion based on the Lindblad formalism, the GIOM is nonperturbative. 
Furthermore, the GIOM formalism provides a simple, flexible framework 
that can be readily implemented for quantum dot circuits with complex connectivity.


\section{Engineering nonreciprocal interactions: Case studies}\label{s:2}

To illustrate the utility of the above general discussion and
Eq. (\ref{eq:smatrix}) on nonreciprocal interaction engineering, 
we study two cases: 
In the three-dot model, engineered dissipation counteract {\it direct} couplings between the primary dots
to induce nonreciprocity.
In contrast, in the four-dot model the two primary dots do not directly couple, 
and two auxiliary dots with engineered dissipation
build up the nonreciprocal coupling between the primary dots.

\subsection{Three-dot system: direct coherent coupling between primary dots}

\begin{figure}[tbh!]
 \centering
\includegraphics[width=0.7\columnwidth] {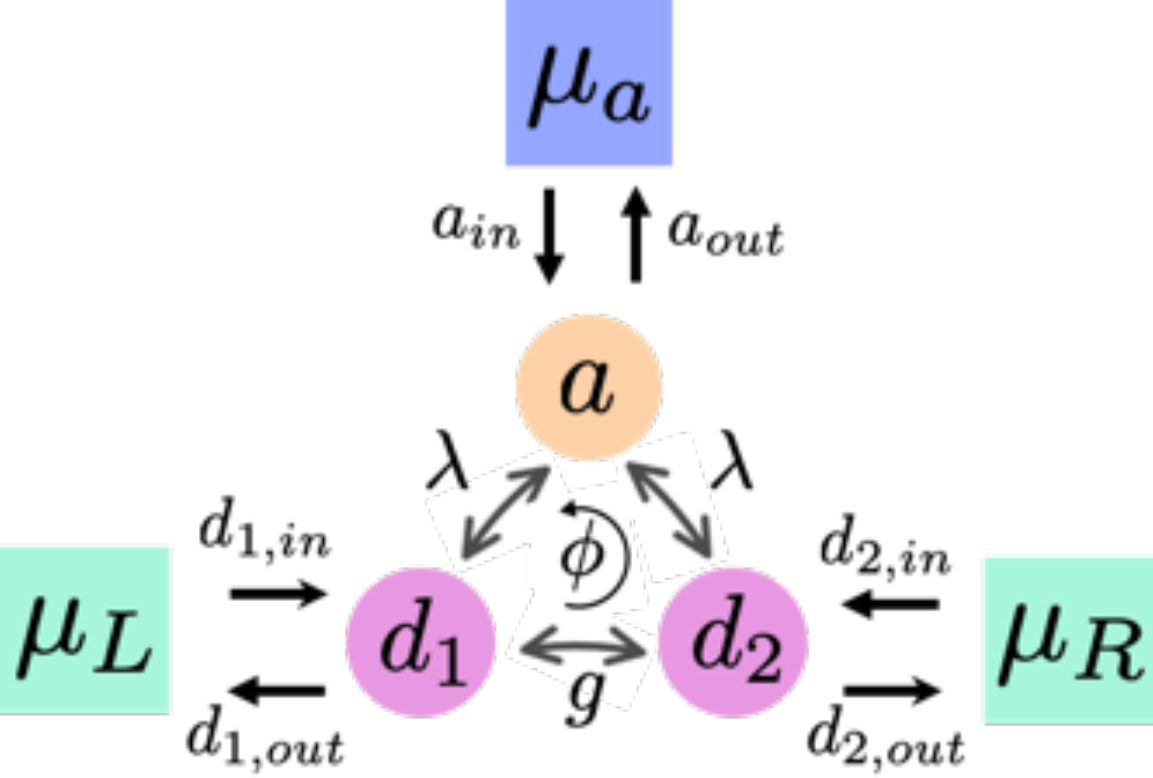} 
\caption{A three-dot system with a loop phase $\phi$. 
Electrons tunnel between primary dots with the coherent tunneling $g$.
$\lambda$ is the tunneling term between primary and auxilary dots.
Each dot is tunnel-coupled to a metallic lead whose chemical potential can be externally controlled.
The auxilary dot $a$ is strongly coupled to a fermionic reservoir with the damping rate $\kappa$.
Nonreciprocal behavior results from the interference between the coherent couplings $g$ and $\lambda$
and the dissipative rate $\kappa$.
}
\label{fig:fig2}
\end{figure}

We first consider the case in which a direct coherent tunneling of electrons between the primary dots is presented. 
For demonstration purposes, we adopt a minimal three-dot system,
with two primary dots (1 and 2) and one damped auxiliary dot (a), as shown in Fig. \ref{fig:fig2}.

\begin{figure}[tbh!]
 \centering
\includegraphics[width=1.05\columnwidth]  {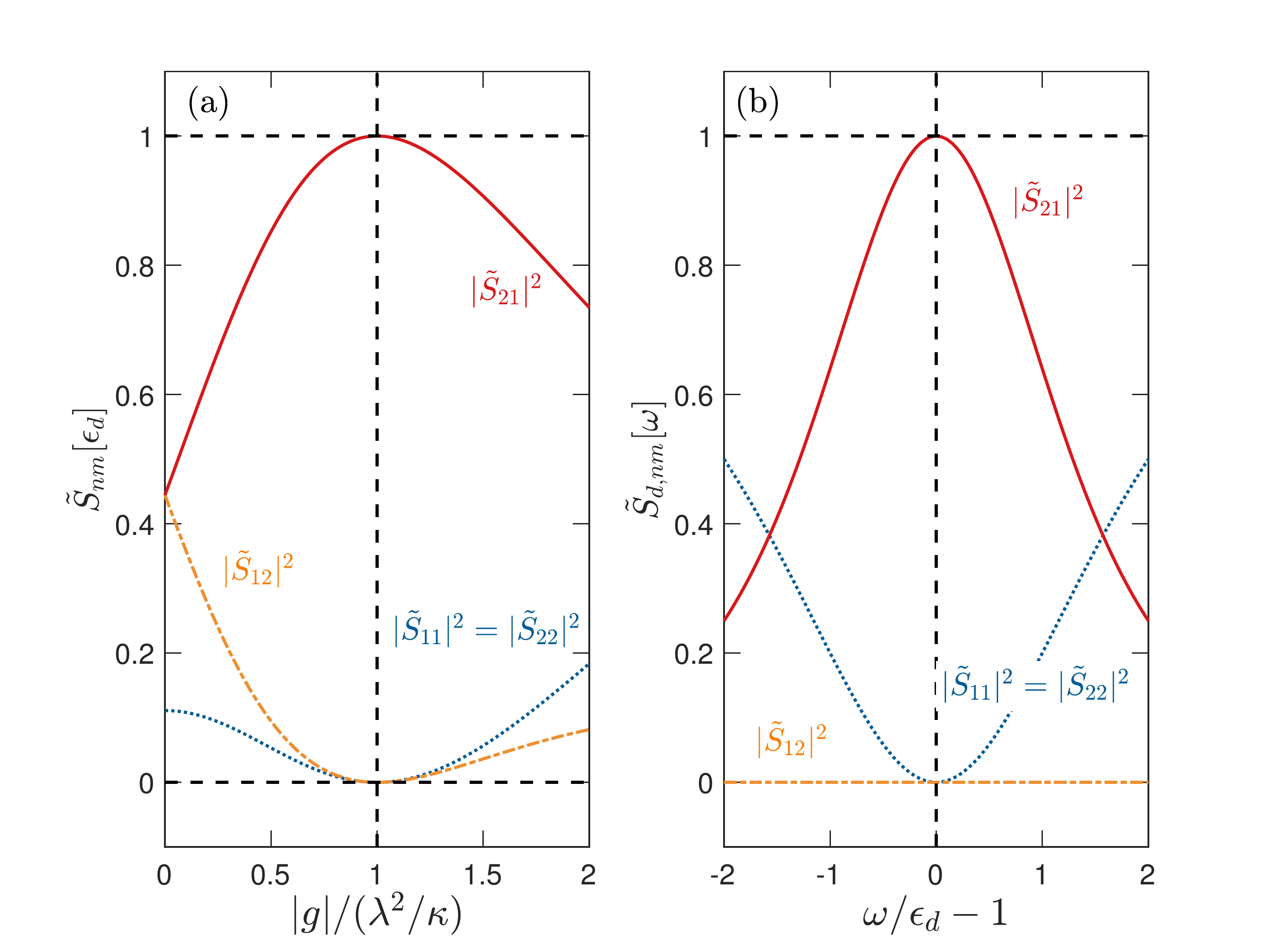} 
\caption{
Nonreciprocal behavior in the three-dot model of Fig. \ref{fig:fig2}.
(a) Scattering matrix elements in Eq. (\ref{eq:s_td}) at frequency $\omega=\epsilon_d$ 
as a function of the tunneling element $g$. 
We have set $\Gamma=\lambda^2/\kappa$. 
The phase of $g$ is fixed such that arg($g/(\lambda^2/\kappa)$)=$\pi/2$. 
(b) Scattering matrix elements in Eq. (\ref{eq:ss}) as a function of frequency $\omega$, 
when the directionality condition as well as as the impedance matching condition are fulfilled. 
In both panels, we set $\Gamma=\epsilon_d=1$. 
}
\label{fig:fig3} 
\end{figure}
The coherent part of the total Hamiltonian 
takes the form
\begin{equation}\label{eq:coh_td}
H_{coh}~=~\sum_{n=1,2}\epsilon_nd_n^{\dagger}d_n+[gd_1^{\dagger}d_2+\lambda a^{\dagger}(d_1+d_2)+\mathrm{H.c.}].
\end{equation}
Here, $d_n$ ($a$) is the annihilation operator for the primary dot $n$ (auxiliary dot $a$), 
$\epsilon_n$ is the corresponding electronic on-site energy for the primary dot $n$ 
(the on-site energy of the auxiliary dot is set at zero, as it plays a negligible role in the large damping limit).
$g$ ($\lambda$) is the coherent coupling between the two primary dots (between primary dots and the auxiliary dot). 

To break the symmetry of the two dots, an external magnetic field is applied. 
As a result, coherent electrons moving in a closed loop pick a phase 
proportional to the magnetic flux through the loop. 
The coherent tunneling $g$ and $\lambda$ therefore become complex.
However, in such a closed loop architecture one can always shift the complex phase of 
$\lambda$ into that of $g$. We thus set $\lambda$ to be real, 
and keep $g$ complex. We also set the levels to be identical,
$\epsilon_n=\epsilon_d$. This ensures that nonreciprocity emerges here due to the combination
of complex coherent coupling and dissipation engineering, and not due to the energetic asymmetry between the dots.

We write down the corresponding HLEs for annihilation operators 
using Eq. (\ref{eq:eom_o}) (explicit time dependence is suppressed):
\bea\label{eq:hle_td}
\dot{d}_1 &=& -i\epsilon_d d_1-\Gamma d_1-igd_2-i\sqrt{2\pi}d_{1,in}-i\lambda a,\nonumber\\
\dot{d}_2 &=& -i\epsilon_d d_2-\Gamma d_2-ig^{\ast}d_1-i\sqrt{2\pi}d_{2,in}-i\lambda a,\nonumber\\
\dot{a} &=& -\kappa a-i\sqrt{2\pi} a_{in}-i\lambda(d_1+d_2),
\eea 
where we have denoted $\dot{A}\equiv dA/dt$, 
$g^{\ast}$ is the complex conjugate of $g$. In the large damping limit, one can adiabatically solve the last HLE \cite{Metelmann.15.PRX}, yielding
\begin{equation}\label{eq:a_hle}
a~=~-i\frac{\sqrt{2\pi}}{\kappa}a_{in}-i\frac{\lambda}{\kappa}(d_1+d_2).
\end{equation}
Inserting it into the HLEs for $d_1$, $d_2$, we find
\bea\label{eq:dd_hle}
\dot{d}_1 &=& -Z d_1-\left(\frac{\lambda^2}{\kappa}+ig\right)d_2-i\sqrt{2\pi}d_{1,in}-\frac{\lambda}{\kappa}\sqrt{2\pi} a_{in},\nonumber\\\dot{d}_2 &=& -Z d_2-\left(\frac{\lambda^2}{\kappa}+ig^{\ast}\right)d_1-i\sqrt{2\pi}d_{2,in}-\frac{\lambda}{\kappa}\sqrt{2\pi} a_{in}
\eea
with $Z=i\epsilon_d+\Gamma+\lambda^2/\kappa$. Note that the coherent coupling involves $g$ in the first line of Eq. (\ref{eq:dd_hle}) and $g^{\ast}$ in the second line.
 Therefore, the two coupling terms may cancel, but only in one of the above two equations. 
Here, we set the directionality condition
\begin{equation}
\label{eq:g_value}
g~=~i\frac{\lambda^2}{\kappa},
\end{equation}
which results in a nonreciprocal interaction: Dot 2 is driven by dot 1 but not vice versa,
as dot 1 is not influenced by dot 2.
This directionality can be reversed by tuning the phase of $g$: 
Let $g=|g|e^{i\phi}$.
 Considering the opposite loop phase $\phi\to-\phi$ 
(this can be achieved by reversing the direction of the applied magnetic field), 
Eq. (\ref{eq:g_value}) turns into $|g|e^{-i\phi}=i\lambda^2/\kappa$ which is just the condition $g^{\ast}=i\lambda^2/\kappa$. 

To identify an optimal configuration, we turn to the scattering matrix by utilizing Eqs. (\ref{eq:a_hle}) and (\ref{eq:dd_hle}) (for simplicity, only the elements of the upper left $2\times 2$ matrix are listed as it corresponds to the scattering matrix for the primary two-dot system):
\bea\label{eq:s_td}
\tilde{S}_{11}[\omega] &=& \frac{[i(\epsilon_d-\omega)+\lambda^2/\kappa]^2-\Gamma^2-(\lambda^2/\kappa+ig)(\lambda^2/\kappa+ig^{\ast})}{[i(\epsilon_d-\omega)+\lambda^2/\kappa+\Gamma]^2-(\lambda^2/\kappa+ig)(\lambda^2/\kappa+ig^{\ast})},\nonumber\\
\tilde{S}_{12}[\omega] &=& \frac{2\Gamma(\lambda^2/\kappa+ig)}{[i(\epsilon_d-\omega)+\lambda^2/\kappa+\Gamma]^2-(\lambda^2/\kappa+ig)(\lambda^2/\kappa+ig^{\ast})},\nonumber\\
\tilde{S}_{21}[\omega] &=& \frac{2\Gamma(\lambda^2/\kappa+ig^{\ast})}{[i(\epsilon_d-\omega)+\lambda^2/\kappa+\Gamma]^2-(\lambda^2/\kappa+ig)(\lambda^2/\kappa+ig^{\ast})},\nonumber\\
\tilde{S}_{22}[\omega] &=& \tilde{S}_{11}[\omega].
\eea
Intriguingly, one can readily obtain the directionality condition Eq. (\ref{eq:g_value}) by setting $\tilde{S}_{12}[\omega]=0$. 

We recall that in the scattering formalism, $\tilde S_{11}[\omega]$ is the reflection amplitude,
of an electron of energy $\omega$ to arrive towards dot 1 from lead $L$, and be scattered back to that same electrode. 
Similarly, $\tilde S_{12}[\omega]$ is the transmission amplitude, for an incoming electron from the $L$ side to be absorbed at dot 1,
and be transmitted to the $R$ terminal through dot 2. The directionality condition requires $\tilde S_{12}[\omega]=0$,
yet allows the reverse transmission process, $\tilde S_{21}[\omega]\neq 0$.

Applying the directionality condition, the scattering matrix elements reduces to ($\tilde{\boldsymbol{S}}_d[\omega]\equiv \tilde{\boldsymbol{S}}[\omega]\Big|_{g=i\lambda^2/\kappa}$)
\bea\label{eq:ss}
\tilde{S}_{d,11}[\omega] &=& \tilde{S}_{d,22}[\omega] = \frac{i(\epsilon_d-\omega)+\lambda^2/\kappa-\Gamma}{i(\epsilon_d-\omega)+\lambda^2/\kappa+\Gamma},\nonumber\\
\tilde{S}_{d,12}[\omega] &=& 0,\nonumber\\
\tilde{S}_{d,21}[\omega] &=& \frac{4\Gamma\lambda^2/\kappa}{[i(\epsilon_d-\omega)+\lambda^2/\kappa+\Gamma]^2}.
\eea
We can further suppress reflections, that is diagonal elements of the scattering matrix. 
In the resonance limit (i.e., $\omega=\epsilon_d$), one immediately finds that $\tilde{S}_{d,11}=\tilde{S}_{d,22}=0$ when tuning
\begin{equation}
\label{eq:gamma_value}
\Gamma~=~\frac{\lambda^2}{\kappa}.
\end{equation}
In analogy to quantum optomechanical systems, we refer to this latter relation
as an impedance matching condition. 
Notably, this condition cannot be inferred from the HLEs Eq. (\ref{eq:hle_td}). 

We point out that $\tilde{S}_{d,21}[\epsilon_d]=1$ when Eq. (\ref{eq:gamma_value}) holds, 
which corresponds to a maximum nonreciprocity between the two primary dots. 
However, this maximum nonreciprocity only occurs for resonant situation,
 with $|\omega-\epsilon_d| \ll \Gamma$, 
while the directionality condition Eq. (\ref{eq:g_value}) holds for all frequencies. 
The behavior of scattering matrix elements as functions of the tunneling element $g$ and the frequency $\omega$ 
is depicted in Fig. \ref{fig:fig3}, from which we clearly observe the optimal conditions,
 as well as parameter ranges where an efficient nonreciprocal interaction can be achieved.
The directionality condition is analyzed in panel a. In panel b, the directionality condition is satisfied,
and we test the resonance condition, which is required for impedance matching.

\subsection{Four-dot system: Indirect coherent coupling between primary dots}
\label{ssec:fd}

We now turn to the situation in which the primary dots are not directly connected. 
To this end, we consider a four-dot system as shown in Fig. \ref{fig:fig5}, 
a minimal model that consists two primary dots and two auxiliary dots. 
The coherent part of the total Hamiltonian in Eq. (\ref{eq:HH}) reads
\bea\label{eq:hcon_fd}
H_{coh} &=& \sum_{n=1,2}\epsilon_nd_n^{\dagger}d_n+\sum_{m=1,2}\delta_ma_m^{\dagger}a_m+\Big[g_{11}d_1a_1^{\dagger}+g_{12}d_1a_2^{\dagger}\nonumber\\
&&+g_{21}d_2a_1^{\dagger}+g_{22}d_2a_2^{\dagger}+H.c.\Big],
\eea
where $d_n$ ($a_m$) is the annihilation operator for the primary dot $n$ (auxiliary dot $m$), $\epsilon_n$ ($\delta_m$) is the corresponding electronic on-site energy for primary dot $n$ (auxiliary dot $m$), $g_{11,12,21,22}$ are the coherent hopping rates between primary and auxiliary dots. We again set $\epsilon_n=\epsilon_d$. Here, the on-site energies for auxiliary dots are important, in particular, we will see below that one
needs to set $\delta_1\neq\delta_2$ so as to ensure the directionality condition to be fulfilled, similar to the situation in quantum optomechanical systems \cite{Bernier.17.NC,Peterson.17.PRX}.
\begin{figure}[tbh!]
 \centering
\includegraphics[width=0.67\columnwidth] {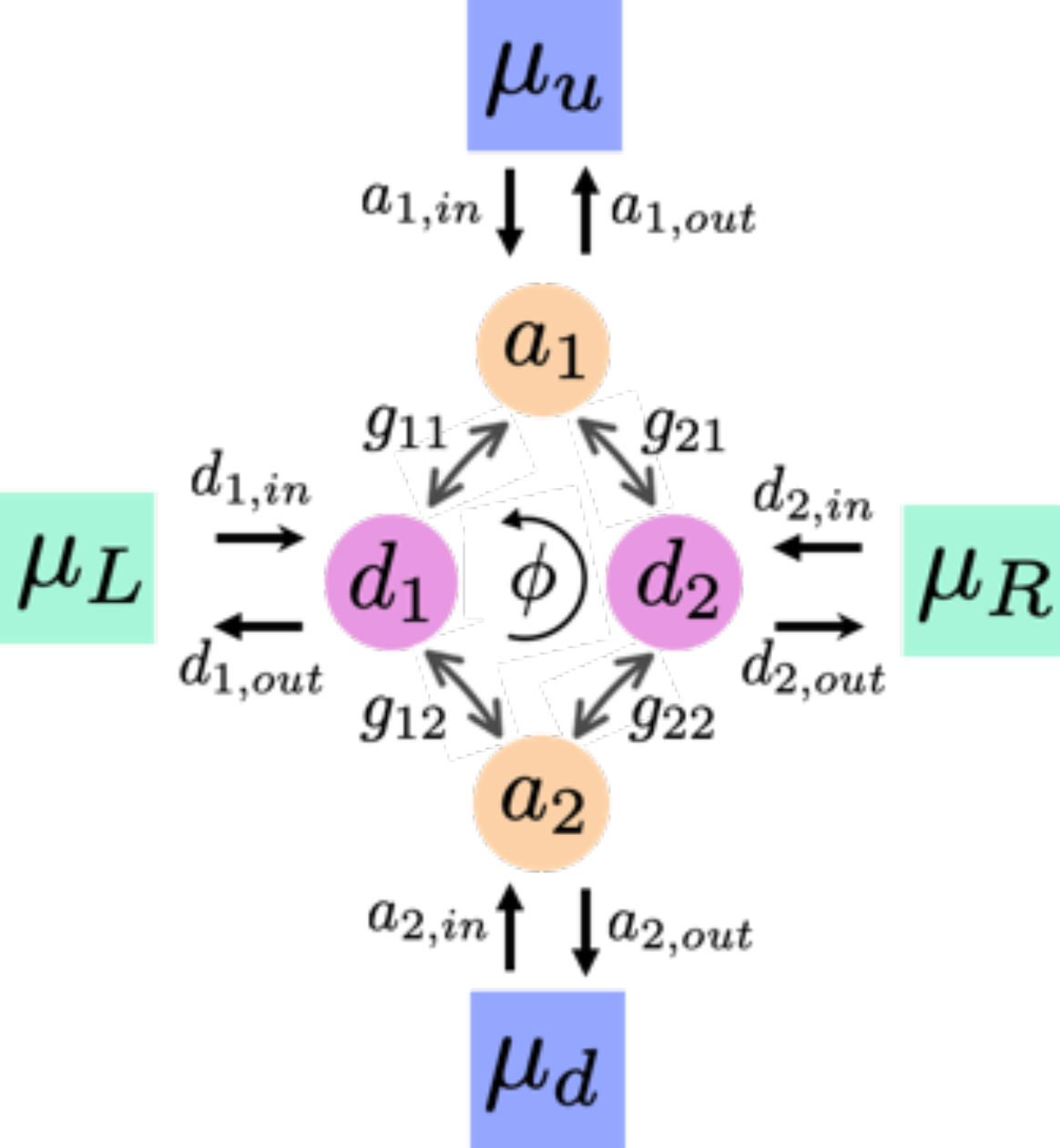} 
\caption{A four-dot system with a loop phase $\phi$ enabling nonreciprocal interaction engineering: Electrons can hop between primary and auxiliary dots with coupling rates $g_{11,12,21,22}$ and each dot is tunnel-coupled to a fermionic reservoir whose chemical potential can be externally controlled.
}
\label{fig:fig4}
\end{figure}

To determine optimal conditions for nonreciprocal interaction, we again resort to the HLEs for annihilation operators by using Eq. (\ref{eq:eom_o}) together with Eq. (\ref{eq:hcon_fd}),
\bea
\dot{d}_1 &=& -i\epsilon_d d_1-\Gamma d_1+ig_{11}^{\ast}a_1+ig_{12}^{\ast}a_2-i\sqrt{2\pi}d_{1,in},\nonumber\\
\dot{d}_2 &=& -i\epsilon_d d_2-\Gamma d_2+ig_{21}^{\ast}a_1+ig_{22}^{\ast}a_2-i\sqrt{2\pi}d_{2,in},\nonumber\\
\dot{a}_1 &=& -i\delta_1a_1-\kappa a_1+ig_{11}d_1+ig_{21}d_2-i\sqrt{2\pi} a_{1,in},\nonumber\\
\dot{a}_2 &=& -i\delta_2a_2-\kappa a_2+ig_{12}d_1+ig_{22}d_2-i\sqrt{2\pi} a_{2,in}.
\eea 
In the large damping limit in which $\kappa$ defines the largest energy scale, one can adiabatically solve HLEs for auxiliary operators, yielding
\bea\label{eq:a_hle_1}
a_1 &=& -i\frac{\sqrt{2\pi}}{\kappa+i\delta_1}a_{1,in}+i\frac{g_{11}}{\kappa+i\delta_1}d_1+i\frac{g_{21}}{\kappa+i\delta_1}d_2,\nonumber\\
a_2 &=& -i\frac{\sqrt{2\pi}}{\kappa+i\delta_2}a_{2,in}+i\frac{g_{12}}{\kappa+i\delta_2}d_1+i\frac{g_{22}}{\kappa+i\delta_2}d_2.
\eea
Inserting them into the HLEs for $d_1$, $d_2$, we find
\bea\label{eq:dd_hle_1}
\dot{d}_1 &=& -\left(i\epsilon_d+\Gamma+\frac{|g_{11}|^2}{\kappa+i\delta_1}+\frac{|g_{12}|^2}{\kappa+i\delta_2}\right) d_1-\Phi d_2\nonumber\\
&&-i\sqrt{2\pi}d_{1,in}+\frac{g_{11}^{\ast}\sqrt{2\pi}}{\kappa+i\delta_1} a_{1,in}+\frac{g_{12}^{\ast}\sqrt{2\pi}}{\kappa+i\delta_2}a_{2,in},\nonumber\\
\dot{d}_2 &=& -\left(i\epsilon_d+\Gamma+\frac{|g_{21}|^2}{\kappa+i\delta_1}+\frac{|g_{22}|^2}{\kappa+i\delta_2}\right) d_2-\Psi d_1\nonumber\\
&&-i\sqrt{2\pi}d_{2,in}+\frac{g_{21}^{\ast}\sqrt{2\pi}}{\kappa+i\delta_1} a_{1,in}+\frac{g_{22}^{\ast}\sqrt{2\pi}}{\kappa+i\delta_2} a_{2,in},
\eea
here, we have denoted $\Phi=\frac{g_{11}^{\ast}g_{21}}{\kappa+i\delta_1}+\frac{g_{12}^{\ast}g_{22}}{\kappa+i\delta_2}$ and $\Psi=\frac{g_{21}^{\ast}g_{11}}{\kappa+i\delta_1}+\frac{g_{22}^{\ast}g_{12}}{\kappa+i\delta_2}$. 

A nonreciprocal interaction between the two primary dots is achieved by letting $\Phi=0$ while keeping $\Psi$ finite, that is, we consider a situation in which primary dot 2 is influenced by primary dot 1 but not vice versa. Without loss of generality, we take all hopping rates to be real except $g_{21}$ with a loop phase $\phi$, $g_{21}=|g_{21}|e^{i\phi}$. To clarify the analytical results, we further assume that each auxiliary dot is equally coupled to both primary dots such that $g_{11}=|g_{21}|=\lambda_1$ and $g_{12}=g_{22}=\lambda_2$. Hence the directionality requirement $\Phi=0$ corresponds to the following loop phase
\begin{equation}\label{eq:phase}
e^{i\phi}~=~-\frac{\kappa+i\delta_1}{\kappa+i\delta_2}\frac{\lambda_2^2}{\lambda_1^2}.
\end{equation}
As can be seen, nonreciprocal interaction is absent if $\delta_1=\delta_2$ since then $e^{i\phi}$ is a real number and both $\Phi$ and $\Psi$ vanish. Similar to the three-dot case, we can reverse the direction of the nonreciprocal interaction at the opposite loop phase.

To gain more insights into the optimal configuration, we examine the scattering matrix obtained from Eqs. (\ref{eq:a_hle_1}) and (\ref{eq:dd_hle_1}) in the Fourier space (for simplicity, only the elements of the upper left $2\times 2$ matrix are listed as it corresponds to the scattering matrix for the primary two-dot system by choosing the basis order $d_1$, $d_2$, $a_1$, $a_2$):
\bea\label{eq:ss_fd}
\tilde{S}_{11}[\omega] &=& \frac{[i(\Delta-\omega)+\Sigma][i(\Delta-\omega)+\Sigma-2\Gamma]-\Phi\Psi}{[i(\Delta-\omega)+\Sigma]^2-\Phi\Psi},\nonumber\\
\tilde{S}_{12}[\omega] &=& \frac{2\Gamma\Phi}{[i(\Delta-\omega)+\Sigma]^2-\Phi\Psi},\nonumber\\
\tilde{S}_{21}[\omega] &=& \frac{2\Gamma\Psi}{[i(\Delta-\omega)+\Sigma]^2-\Phi\Psi},\nonumber\\
\tilde{S}_{22}[\omega] &=& \tilde{S}_{11}[\omega],
\eea
where we have denoted $\Delta=\epsilon_d-\frac{\lambda_1^2\delta_1}{\kappa^2+\delta_1^2}-\frac{\lambda_2^2\delta_2}{\kappa^2+\delta_2^2}$, $\Sigma=\Gamma+\frac{\lambda_1^2\kappa}{\kappa^2+\delta_1^2}+\frac{\lambda_2^2\kappa}{\kappa^2+\delta_2^2}$. We readily obtain the directionality condition Eq. (\ref{eq:phase}) by letting $\tilde{S}_{12}[\omega]=0$. 

Considering the {\it on-resonance} situation with $\omega=\Delta$ and impedance matching the system with $\Sigma=2\Gamma$ or equivalently
\begin{equation}\label{eq:gamma_value_fd}
\Gamma~=~\frac{\lambda_1^2\kappa}{\kappa^2+\delta_1^2}+\frac{\lambda_2^2\kappa}{\kappa^2+\delta_2^2},
\end{equation}
we obtain the following scattering matrix elements
\begin{equation}
\tilde{S}_{11}[\Delta]=\tilde{S}_{22}[\Delta]=\tilde{S}_{12}[\Delta]=0,~\tilde{S}_{21}[\Delta]=\frac{\Psi}{2\Gamma}
\end{equation}
under the directionality condition $\Phi=0$ [cf. Eq. (\ref{eq:phase})]. The expression for $\tilde{S}_{21}[\Delta]$ further highlights the necessity of $\delta_1\neq\delta_2$ for the appearance of nonreciprocal interactions as $\Psi\propto \delta_1-\delta_2$ when the directionality condition is fulfilled. 

\begin{figure}[tbh!]
 \centering
\includegraphics[width=1.1\columnwidth] {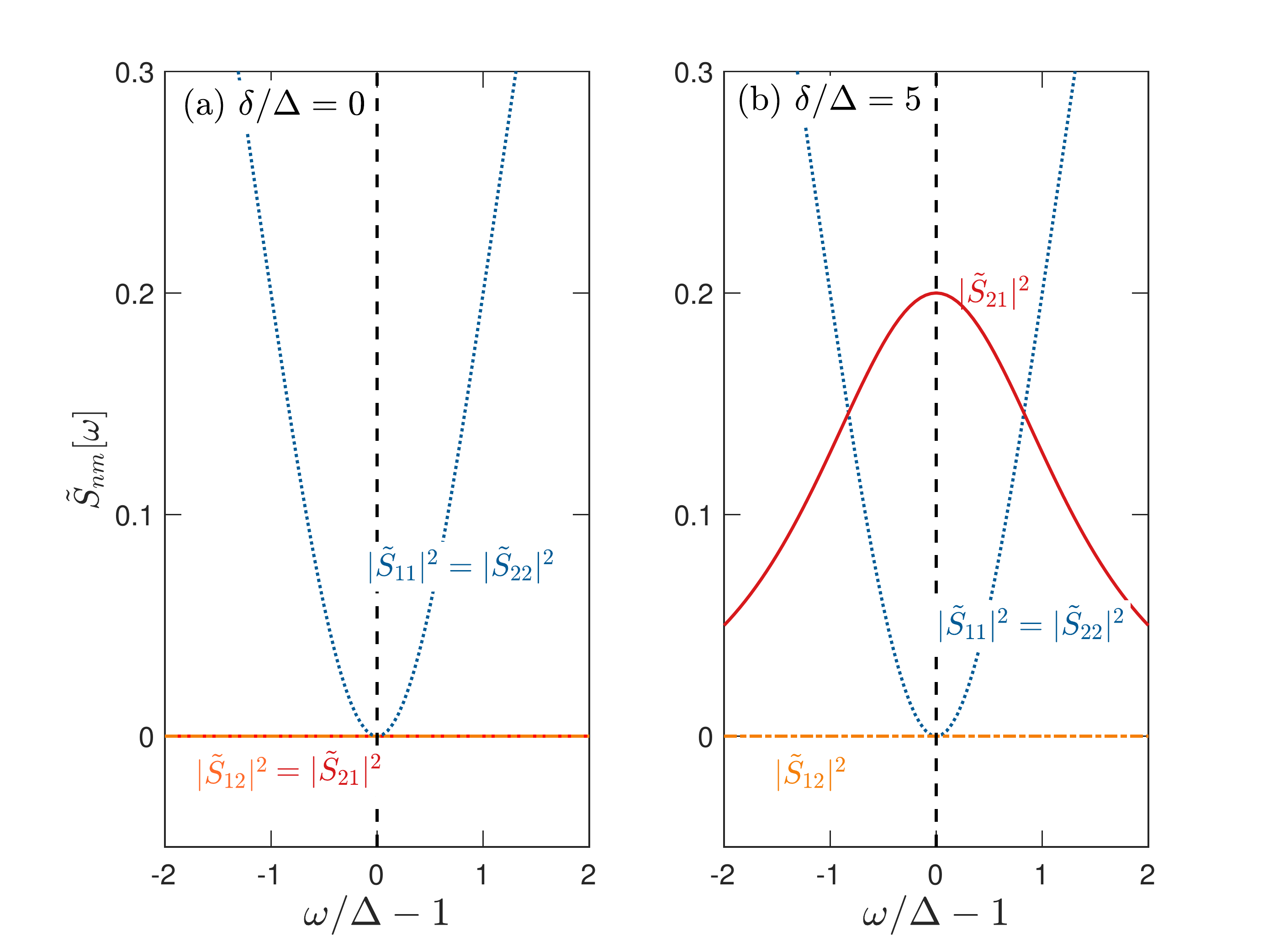} 
\caption{Scattering matrix elements in Eq. (\ref{eq:ss_fd}) as a function of frequency $\omega$ for (a) $\delta/\Delta=0$ and (b) $\delta/\Delta=5$, when the directionality condition as well as as the impedance matching condition are fulfilled. We have set $\Gamma/\Delta=1$ and $\kappa/\Delta=10$.
}
\label{fig:fig5}
\end{figure}
The behavior of the scattering matrix elements as a function of frequency is depicted in Fig. \ref{fig:fig5}. 
In the calculation, we choose $\lambda_1=\lambda_2=\lambda$, $\delta_1=-\delta_2=\delta$ such that $\Psi=2\Gamma\frac{i\delta}{\kappa+i\delta}$ when the directionality condition is applied. In Fig. \ref{fig:fig5} (a), we show that there is simply no tunneling between two primary dots when $\delta_1=\delta_2=0$; the transmissions $|\tilde{S}_{12}[\omega]|^2=|\tilde{S}_{21}[\omega]|^2=0$ indicate that electrons from primary dots are all absorbed by auxiliary dots. Trivially, there is also no nonreciprocity in this case.
When $\delta_1\neq\delta_2$ as depicted in Fig. \ref{fig:fig5} (b), we can achieve nonreciprocity. However, we should point out that here the scattering matrix cannot be optimized to the ideal case in which $|\tilde{S}_{21}[\Delta]|^2=1$, in contrast to the  three-dot system. 
In principle. $|\tilde{S}_{21}[\Delta]|^2=\delta^2/(\kappa^2+\delta^2)$ approaches 1  when $\delta\gg\kappa$, but this is at odds with our assumption that $\kappa$ corresponds to the largest energy scale in the problem. 

\section{Nonequilibrium charge transport}\label{s:3}

So far, we discussed nonreciprocity for charge transport by studying the properties of the 
scattering matrix, with $\omega$ as the energy of an incoming electron, [see Eq. (\ref{eq:smatrix})].
However, unlike in optomechanical systems, where it is sufficient to analyze the behavior at a particular frequency \cite{Metelmann.15.PRX}, 
for electron transport the observable is the electrical current, integrated over electron transmission within the bias window.
While the directionality condition does not depend on frequency, 
impedance matching relies on the resonance condition to eliminate reflections,
see the discussion of Eq. (\ref{eq:gamma_value}). 
As such, it is clear that we cannot exactly satisfy  the zero reflection condition for the net charge current,
and it is important to analyze the extent of nonreciprocity in the behavior of the current at finite bias voltage. 

In this Section, we analyze signatures of nonreciprocal interaction from the perspective of charge transport,
 when chemical potentials of the attached primary leads are tuned to be different. We require that chemical potentials of auxiliary leads are much smaller than those of primary leads so as to ensure that electrons and noise from the auxiliary leads do not transmit to the primary system,
 a feature favored by quantum signal processing applications. We adopt a convention that forward voltage bias corresponds to $\mu_L>\mu_R$. 
As such, nonreciprocity for charge current corresponds to observing $J_L(-V)\neq J_R(V)$ ($V>0$), 
that is, the current towards the left metal (from the right electrode, since the auxiliary electrodes cannot feed in electrons) in the reverse (negative) bias regime is different from the current reaching the right electrode from the left one in the forward bias regime. 
 
Observables of interests are the steady state charge currents out of left (L) and right (R) lead (see Figs. \ref{fig:fig2} and \ref{fig:fig4} for illustrations). In the context of GIOM, they have the following formally-exact definitions in the Heisenberg picture \cite{Liu.20.A} (it is sufficient to work with input fields due to input-output relations Eq. (\ref{eq:in_out}))
\bea\label{eq:J_definition}
J_L &=& 2\left(\sqrt{2\pi}\mathrm{Im}\langle d_{1}^{\dagger}d_{1,in}\rangle-\Gamma\langle d_{1}^{\dagger}d_{1}\rangle\right),\nonumber\\
J_R &=& 2\left(\sqrt{2\pi}\mathrm{Im}\langle d_{2}^{\dagger}d_{2,in}\rangle-\Gamma\langle d_{2}^{\dagger}d_{2}\rangle\right).
\eea
Here, ``Im" refers to an imaginary part, the ensemble averages are performed with respect to an initial factorized state for dots and leads. Specifically, we assume that the metallic leads are initially in their thermal equilibrium states characterized by the Fermi-Dirac distribution $n_F^v(\epsilon)=\{\exp[(\epsilon-\mu_v)/T]+1\}^{-1}$ with $\mu_v$ the corresponding chemical potential and $T$ the temperature. With the initial thermal equilibrium states, statistics for input fields can be defined (see appendix \ref{a:1} for details). As the GIOM conserves the total charge in the metal-dots system \cite{Liu.20.A}, the current flowing into the auxiliary leads can be inferred from charge conservation.

Unless otherwise stated, we assume in the following charge current calculations that the system has been tuned to satisfy both the directionality condition and the impedance matching condition [cf. Eqs. (\ref{eq:g_value}) and (\ref{eq:gamma_value}) for the three-dot system, Eqs. (\ref{eq:phase}) and (\ref{eq:gamma_value_fd}) for the four-dot system].

\subsection{Three-dot system}
To obtain steady state charge currents for the three-dot system, we should first solve Eq. (\ref{eq:dd_hle}) in the steady state limit, and then insert those solutions into definitions Eq. (\ref{eq:J_definition}). A straightforward evaluation leads to (details can be found in appendix \ref{a:2})
\bea
J_L &=& \int\,\frac{d\epsilon}{2\pi}\frac{4\Gamma^2}{4\Gamma^2+(\epsilon_d-\epsilon)^2}\Big[n_F^L(\epsilon)-n_F^A(\epsilon)\Big],\label{eq:JL}\\
J_R &=& \int\,\frac{d\epsilon}{2\pi}\frac{4\Gamma^2}{4\Gamma^2+(\epsilon_d-\epsilon)^2}\Big[n_F^R(\epsilon)-n_F^A(\epsilon)\Big]\nonumber\\
&&-\int\,\frac{d\epsilon}{2\pi}\frac{16\Gamma^4}{[4\Gamma^2+(\epsilon_d-\epsilon)^2]^2}\Big[n_F^L(\epsilon)-n_F^A(\epsilon)\Big].\label{eq:JR}
\eea
As can be seen, $J_L$ is independent of the right lead influence due to the nonreciprocal interaction. On the contrary, $J_R$ has contributions from both the left lead and the auxiliary reservoir. Noting that we consider the parameter regime of $\mu_a\ll\min(\mu_L,\mu_R)$, which ensures that no electrons as well as noise from the auxiliary reservoir transmit to the system. 
Therefore, according to Eqs. (\ref{eq:JL}) and (\ref{eq:JR}), $J_L$ is always positive (electron leave the $L$ lead), irrespective of the direction of voltage bias between the two primary dots, while $J_R$ can become negative, thereby allowing electrons from the left lead to reach into the right one.

%
\begin{figure}[tbh!]
 \centering
\includegraphics[width=1\columnwidth] {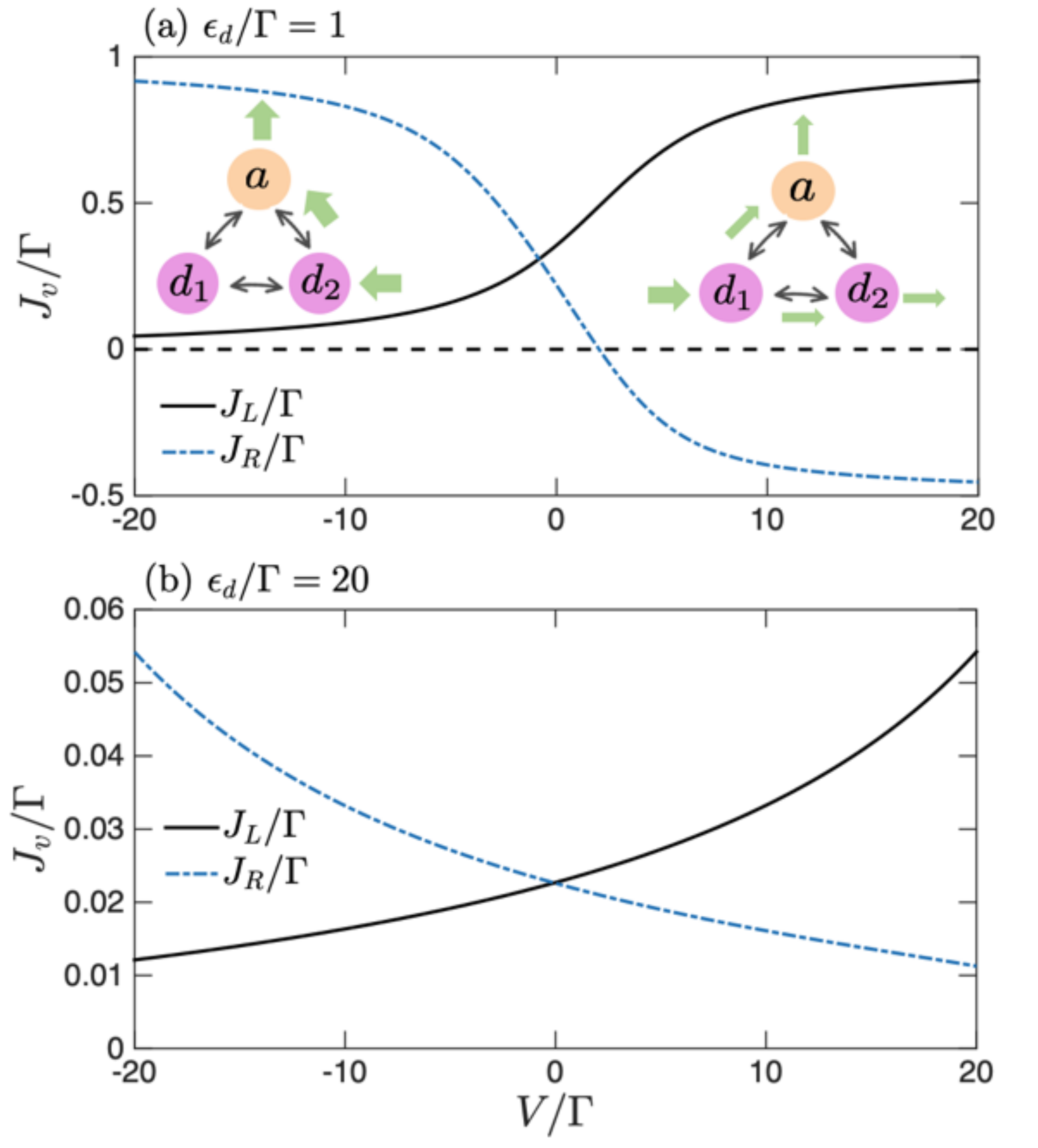}  
\caption{Charge currents out of left lead ($J_L$, solid line) and right lead ($J_R$, dashed-dotted line) as a function of voltage bias $V$ for (a) resonant transport with $\epsilon_d/\Gamma=1$ and (b) off-resonant transport with $\epsilon_d/\Gamma=20$. 
The left and right insets in (a) show ideal charge transfer pathways (as indicated by green arrows) in the large reverse and forward bias regimes, respectively. We set $\mu_L=V/2$, $\mu_R=-V/2$, $\mu_a/\Gamma=-50$ and $T/\Gamma=0.5$.
Other parameters are selected so as to satisfy Eqs. (\ref{eq:g_value}) and  (\ref{eq:gamma_value}). 
}
\label{fig:fig6}
\end{figure}
%
The behavior of the steady state charge currents, $J_L$ and $J_R$, are shown in Fig. \ref{fig:fig6}. 
We consider both resonant and off-resonant regimes, noting that the impedance matching condition, Eq. (\ref{eq:gamma_value}) relies on resonant transport behavior (see also Fig. \ref{fig:fig3} (b)). As can be seen from the Fig. \ref{fig:fig6} (a) for the resonant transport case, in the negative bias regime with $\mu_L<\mu_R$, electrons out of the right lead are routed into the auxiliary dot without flowing into the left lead (otherwise $J_L$ should become negative). While in the case of forward (positive) bias, electrons out of the left lead follow two transport pathways: One part is transmitted into the right lead and the other part is absorbed by the auxiliary dot. The corresponding electron transfer pathways in the reverse and forward bias regimes are highlighted in the insets of Fig. \ref{fig:fig6} (a).
 Overall, the intrinsic dissipation of the auxiliary dot, which takes electrons out of the system regardless of the transmission direction, is an essential ingredient for the nonreciprocal interaction engineering reported here.
 This figure illustrates one of the main results of this work, that we observe nonreciprocity for the integrated charge current, 
  not just the scattering matrix: At high positive voltage, current flows from the left to the right lead. However, there is no net current from the right metal to the left one when we reverse the voltage bias, $J_L(-V) \neq J_R(V)$.
  
In Fig. \ref{fig:fig6} (b), we further consider the off-resonant charge transport case. In this regime, the second term on the right-hand-side (RHS) of Eq. (\ref{eq:JR}) becomes negligible compared to the first term, hence $J_R$ and $J_L$ become symmetric under the transformation $L\leftrightarrow R$, thereby indicating that there is no nonreciprocal behavior between the two primary dots: Injected electrons from both sides are all absorbed by the auxiliary dot towards its reservoirs. Those features are clearly visible from Fig. \ref{fig:fig6} (b): Both $J_L$ and $J_R$ are positive in the whole voltage bias regime and become almost symmetric about $V=0$. 

%
\begin{figure}[tbh!]
 \centering
\includegraphics[width=1\columnwidth] {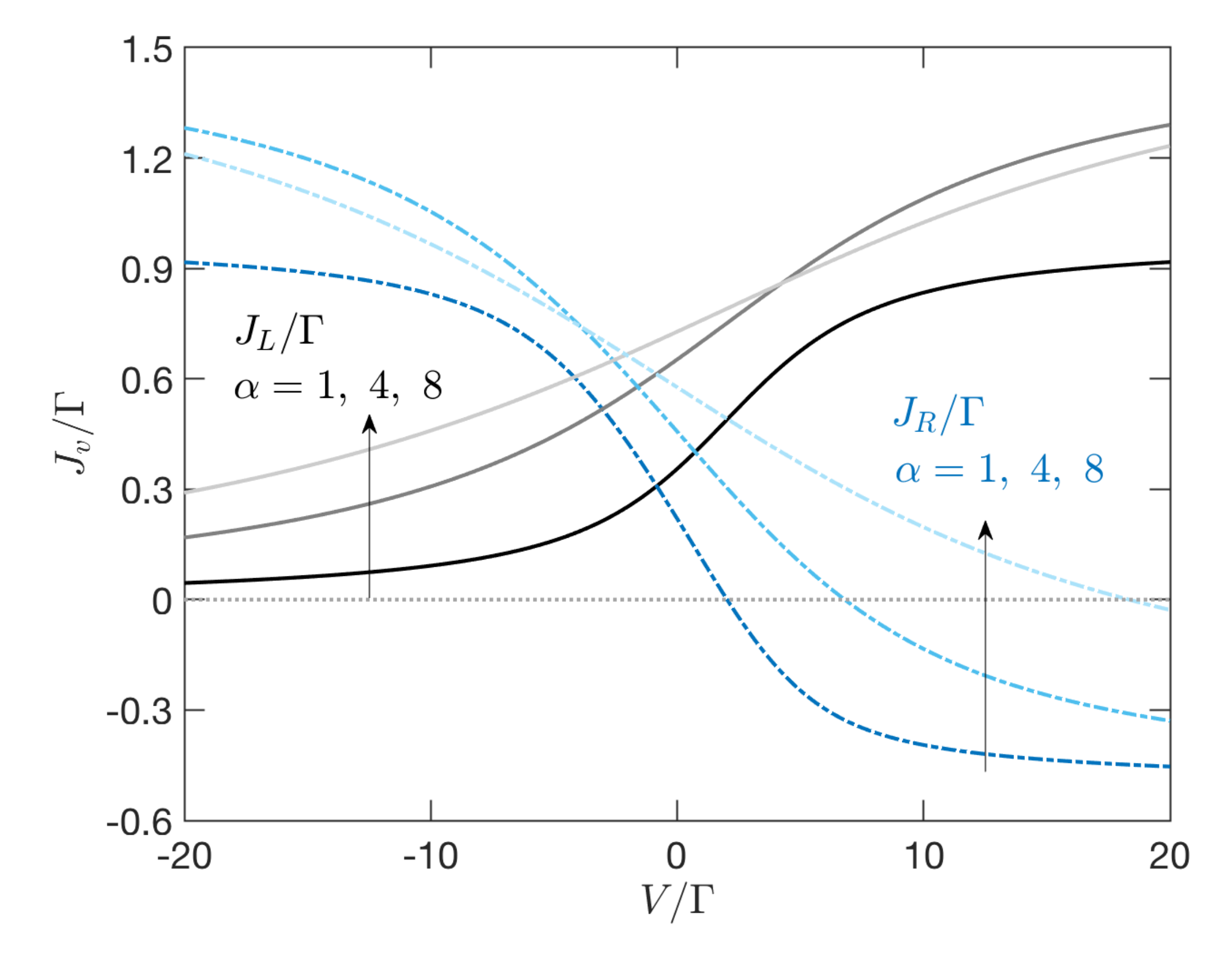} 
\caption{Charge currents out of left lead ($J_L$, solid lines) and right lead ($J_R$, dashed-dotted lines) as a function of voltage bias $V$ for the ratios $\alpha=\lambda^2/(\kappa\Gamma)=1,~4,~8$, relaxing the impedance condition.
 The directionality condition is imposed. Other parameters are the same with Fig. \ref{fig:fig6} (a).
}
\label{fig:fig7}
\end{figure}

\begin{figure}[tbh!]
 \centering
\includegraphics[width=1\columnwidth] {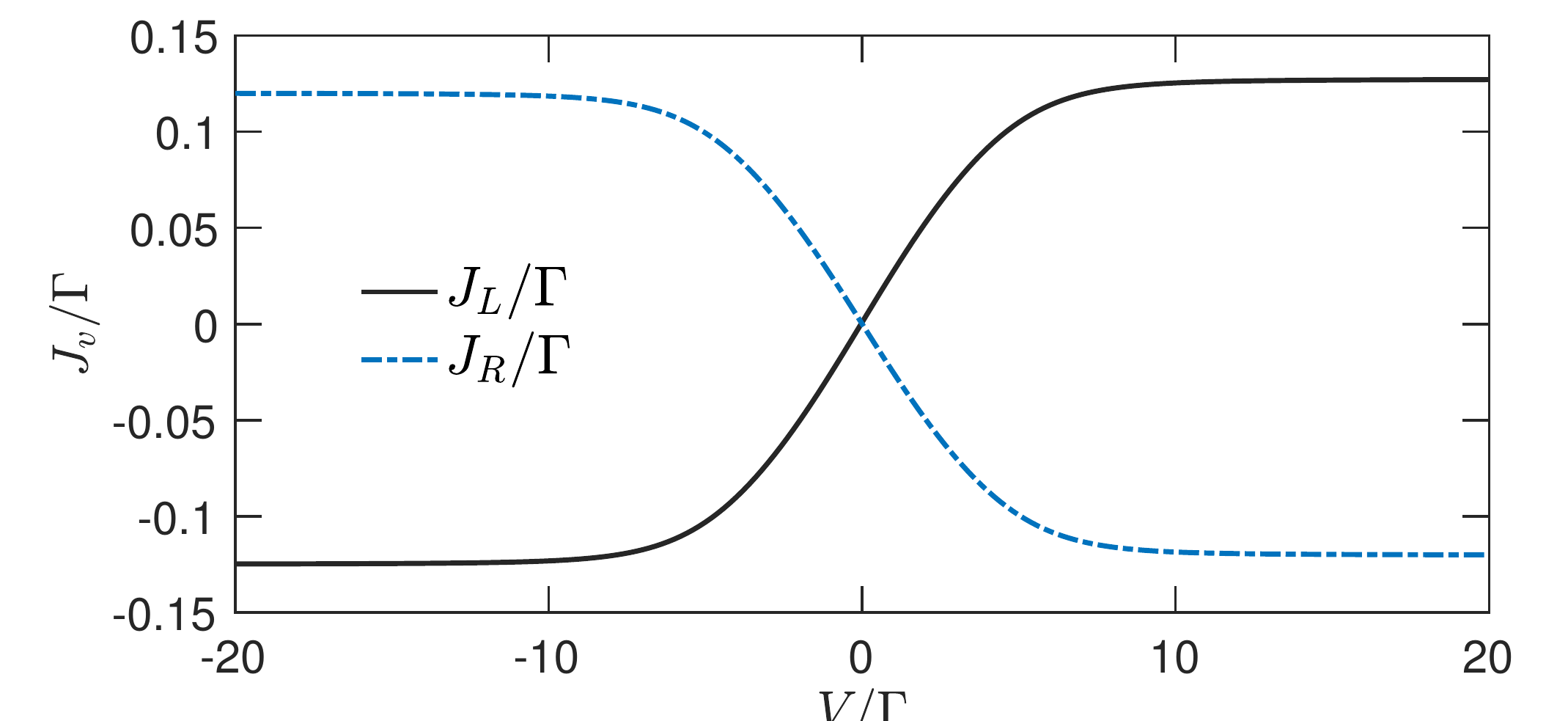} 
\caption{
Charge currents with {\it reciprocal} tunneling elements in a three-dot system. 
The currents, out of left lead ($J_L$, solid line) and right lead ($J_R$, dashed-dotted line) are plotted
as a function of the voltage bias $V$, and obtained from the Landauer-B\"uttiker formula. 
We set $\epsilon_d/\Gamma=1$, $|g|/\Gamma=1$ with a fixed loop phase $\phi=\pi$, 
$\lambda/\Gamma=1$ and $\kappa/\Gamma=100$. 
Other parameters are the same with Fig. \ref{fig:fig6}.
}
\label{fig:fig8}
\end{figure}

Clearly, the directionality condition is the basis for nonreciprocal interactions. To highlight that the impedance matching condition Eq. (\ref{eq:gamma_value}) is also crucial in nonreciprocal interaction engineering, we show resonant charge current results without imposing the condition  (\ref{eq:gamma_value}) in Fig. \ref{fig:fig7}, based on the following general expressions [cf. Eqs. (\ref{eq:jl_b}) and (\ref{eq:jr_b}) in Appendix \ref{a:2}]:
\bea
J_L &=& \int\,\frac{d\epsilon}{2\pi}\frac{4\Gamma^2\alpha}{\Gamma^2(1+\alpha)^2+(\epsilon_d-\epsilon)^2}[n_F^L(\epsilon)-n_F^A(\epsilon)],\nonumber\\
J_R &=& \int\,\frac{d\epsilon}{2\pi}\frac{4\Gamma\lambda^2/\kappa}{\Gamma^2(1+\alpha)^2+(\epsilon_d-\epsilon)^2}\Big[n_F^R(\epsilon)-n_F^A(\epsilon)\Big]\nonumber\\
&&-\int\,\frac{d\epsilon}{2\pi}\frac{16\Gamma^4\alpha^2}{[\Gamma^2(1+\alpha)^2+(\epsilon_d-\epsilon)^2]^2}\Big[n_F^L(\epsilon)- n_F^A(\epsilon)\Big],\nonumber\\
\eea
Here, we have introduced the dimensionless ratio $\alpha=\lambda^2/(\kappa\Gamma)$. 
Compared with Eqs. (\ref{eq:JL}) and (\ref{eq:JR}), we note that the transmission functions are modified when $\alpha\neq1$,
but that the overall transport trends are the same. 
 Hence, we still have a nonreciprocal behavior as can be seen from this figure, but the contrast between the forward and reverse bias regimes becomes less and less significant as $\alpha$ increases. Particularly, the small magnitude of $J_R$ for $\alpha=8$ in the forward regime indicates that one needs relatively large voltage values to efficiently pump electron from the left to the right lead against reflection at the primary dot 1.

For comparison, it is also useful to explore charge transport behavior with {\it reciprocal} couplings, that is, from the perspective of coherent transport as the whole subsystem (primary and auxiliary dots) described by $H_{coh}$ is fully coherent. 
Noticing that the charge current obtained from the GIOM is equivalent to that of the Landauer-B\"uttiker (LB) theory in the coherent limit \cite{Liu.20.A}, we can also directly adopt the multiterminal LB expression \cite{Landauer.57.IBM,Buttiker.88.PRB} for charge current:
\begin{equation}\label{eq:LB}
J_v~=~\sum_{v'\neq v}\int\,\frac{d\epsilon}{2\pi}[\mathcal{T}_{vv'}(\epsilon,\phi)n_F^v(\epsilon)-\mathcal{T}_{v'v}(\epsilon,\phi)n_F^{v'}(\epsilon)],
\end{equation}
where the transmission functions $\mathcal{T}_{vv'}(\epsilon,\phi)=\mathrm{Tr}[\boldsymbol{\Gamma}_v\boldsymbol{G}^r(\epsilon,\phi)\boldsymbol{\Gamma}_{v'}\boldsymbol{G}^a(\epsilon,\phi)]$ with $\boldsymbol{\Gamma}_v$ the dot-lead coupling matrix, $[\boldsymbol{G}^r(\epsilon,\phi)]^{-1}=\epsilon\boldsymbol{I}-\boldsymbol{H}_{coh}(\phi)+i\sum_{v}\boldsymbol{\Gamma}_v=[\boldsymbol{G}^{a,\ast}(\epsilon,\phi)]^{-1}$, here $\boldsymbol{H}_{coh}(\phi)$ is the matrix form for $H_{coh}(\phi)$ with a loop phase $\phi$. $\boldsymbol{I}$ is the identity matrix. 

For the three-dot system, we have $\boldsymbol{\Gamma}_L=\mathrm{diag}(\Gamma,0,0)$, $\boldsymbol{\Gamma}_R=\mathrm{diag}(0,\Gamma,0)$ and $\boldsymbol{\Gamma}_a=\mathrm{diag}(0,0,\kappa)$ and 
\begin{equation}
\boldsymbol{H}_{coh}(\phi)~=~\left(
\begin{array}{ccc}
\epsilon_d & |g|e^{i\phi} & \lambda\\
|g|e^{-i\phi} & \epsilon_d & \lambda\\
\lambda & \lambda & 0
\end{array}
\right)
\end{equation}
A direct calculation leads to 
\bea\label{eq:tt_td}
\mathcal{T}_{LR}(\epsilon,\phi) &=& \mathcal{T}_{RL}(\epsilon,-\phi)~=~\frac{\Gamma^2\left||g|e^{i\phi}(\epsilon+i\kappa)+\lambda^2\right|^2}{\mathbb{D}(\epsilon)},\nonumber\\
\mathcal{T}_{La}(\epsilon,\phi) &=& \mathcal{T}_{aL}(\epsilon,-\phi)~=~\frac{\kappa\Gamma\lambda^2\left|\epsilon_d-|g|e^{i\phi}-\epsilon-i\Gamma\right|^2}{\mathbb{D}(\epsilon)},\nonumber\\
\mathcal{T}_{Ra}(\epsilon,\phi) &=& \mathcal{T}_{aR}(\epsilon,-\phi)~=~\frac{\kappa\Gamma\lambda^2\left|\epsilon_d-|g|e^{i\phi}-\epsilon+i\Gamma\right|^2}{\mathbb{D}(\epsilon)},\nonumber\\
\eea
here, $\mathbb{D}(\epsilon)=\Big|2\lambda^2(\epsilon_d+i\Gamma-\mathrm{Re}[g]-\epsilon)+(\epsilon-i\kappa)\Big((\epsilon_d+i\Gamma-\epsilon)^2-|g|^2\Big)\Big|^2$. 
Interestingly, one can infer the directionality condition Eq. (\ref{eq:g_value}) by letting $\mathcal{T}_{LR}(\epsilon)=0$ in the large damping limit of $\kappa\gg\epsilon$. 
As a comparison, we should point out that it is impossible to engineer a directionality in a two-dot system from the LB expression as we always have 
$\mathcal{T}_{LR}(\epsilon)=\mathcal{T}_{RL}(\epsilon)=(\Gamma^2|g|^2)/(|(\epsilon_d-\epsilon+i\Gamma)^2-|g|^2|^2)$, thereby highlighting the important role of a damped auxiliary dot in engineering nonreciprocal interactions.

 In Fig. \ref{fig:fig8} we present the charge current when $e^{i\phi}=-1$, which leads to $g=g^*$ and thus to {\it reciprocal} tunneling 
 in Eq. (\ref{eq:dd_hle}).
 Calculations are done with the LB expression [cf. Eq. (\ref{eq:LB}) together with Eq. (\ref{eq:tt_td})]. 
It is evident that the nonreciprocal behavior is absent, with $J_L(V) = J_R(-V)$, as compared to Fig. \ref{fig:fig6} (a). Interestingly, we note that the charge currents in the present case, without directionality condition, are about one order of magnitude smaller than currents in the optimal configuration shown in Fig. \ref{fig:fig6} (a). This is because the latter is built upon a single dot transmission function, $4\Gamma^2/[4\Gamma^2+(\epsilon_d-\epsilon)^2]$, and its square according to Eqs. (\ref{eq:JL}) and (\ref{eq:JR}). Hence the nonreciprocal interaction engineering not only induce unidirectional transport, but also enhances the magnitudes of charge currents.

\subsection{Four-dot system}
We now turn to demonstrate that the charge current  in a nonequilibrium four-dot system also demonstrates nonreciprocity. 
For simplicity, we take $\lambda_1=\lambda_2=\lambda$, $\delta_1=-\delta_2=\delta$ in the following. 
Steady state charge currents for the four-dot system can be obtained in a similar manner as in the three-dot case (details can be found in 
Appendix \ref{a:3})
\bea
J_L &=& \int\,\frac{d\epsilon}{2\pi}\frac{2\Gamma^2}{4\Gamma^2+(\epsilon_d-\epsilon)^2}\Big[n_F^L(\epsilon)-n_F^u(\epsilon)\Big]\nonumber\\
&&+\int\,\frac{d\epsilon}{2\pi}\frac{2\Gamma^2}{4\Gamma^2+(\epsilon_d-\epsilon)^2}\Big[n_F^L(\epsilon)-n_F^d(\epsilon)\Big],\label{eq:JL_fd}\\
J_R &=& \int\,\frac{d\epsilon}{2\pi}\frac{2\Gamma^2}{4\Gamma^2+(\epsilon_d-\epsilon)^2}\Big[n_F^R(\epsilon)-n_F^u(\epsilon)\Big]\nonumber\\
&&+\int\,\frac{d\epsilon}{2\pi}\frac{2\Gamma^2}{4\Gamma^2+(\epsilon_d-\epsilon)^2}\Big[n_F^R(\epsilon)-n_F^d(\epsilon)\Big]\nonumber\\
&&-\frac{\delta^2}{\kappa^2+\delta^2}\int\,\frac{d\epsilon}{2\pi}\frac{4\Gamma^4}{[4\Gamma^2+(\epsilon_d-\epsilon)^2]^2}\Big[n_F^L(\epsilon)-n_F^u(\epsilon)\Big]\nonumber\\
&&-\frac{\delta^2}{\kappa^2+\delta^2}\int\,\frac{d\epsilon}{2\pi}\frac{4\Gamma^4}{[4\Gamma^2+(\epsilon_d-\epsilon)^2]^2}\Big[n_F^L(\epsilon)-n_F^d(\epsilon)\Big]\nonumber\\
&&-\frac{4\Gamma^3\delta\kappa}{\kappa^2+\delta^2}\int\,\frac{d\epsilon}{2\pi}\frac{\epsilon_d-\epsilon}{[4\Gamma^2+(\epsilon_d-\epsilon)^2]^2}\Big[n_F^u(\epsilon)-n_F^d(\epsilon)\Big].\nonumber\\
\label{eq:JR_fd}
\eea
Note that if we take $\delta=0$, $J_L$ and $J_R$ become symmetric with $L\leftrightarrow R$, implying the absence of nonreciprocal interactions as we pointed out in Sec. \ref{ssec:fd}. 

\begin{figure}[tbh!]
 \centering
\includegraphics[width=1\columnwidth] {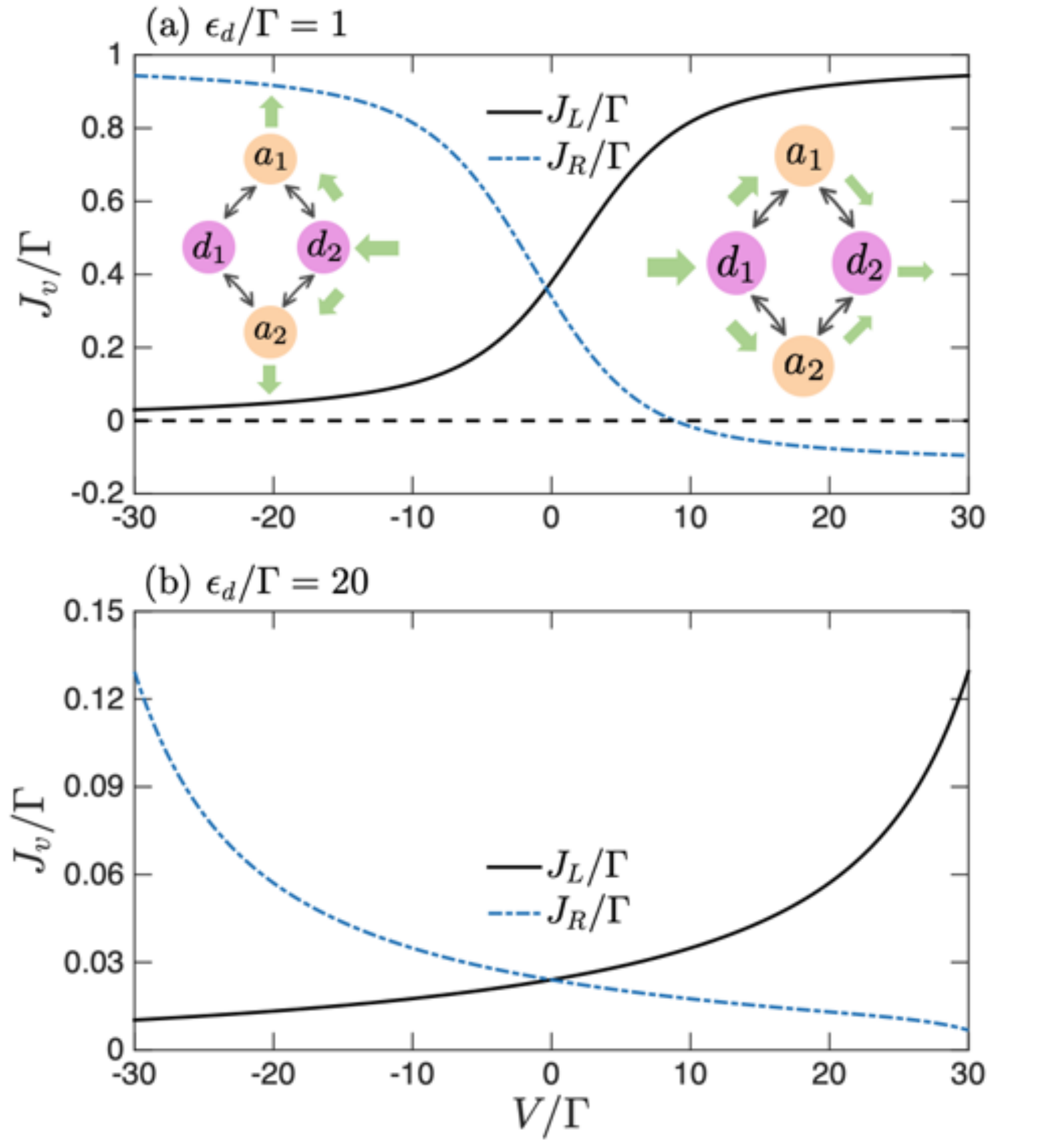} 
\caption{Charge currents out of left lead ($J_L$, solid line) and right lead ($J_R$, dashed-dotted line) as a function of voltage bias $V$ for (a) resonant transport with $\epsilon_d/\Gamma=1$ and (b) off-resonant transport with $\epsilon_d/\Gamma=20$. Left and right insets in (a) show ideal charge transfer pathways (as indicated by green arrows) in the large reverse and forward bias regime, respectively. We set $\mu_L=V/2$, $\mu_R=-V/2$, $\mu_u/\Gamma=\mu_d/\Gamma=-60$, $\epsilon_d/\Gamma=1$, $\delta/\Gamma=30$, $\kappa/\Gamma=30$ and $T/\Gamma=1$.
}
\label{fig:fig9}
\end{figure}

The behavior of the charge currents based on Eqs. (\ref{eq:JL_fd}) and (\ref{eq:JR_fd}) is shown in Fig. \ref{fig:fig9}. In the calculations, we consider the parameter regime of $\mu_u, \mu_d\ll\min(\mu_L,\mu_R)$, which ensures that electrons from the auxiliary reservoir do not enter the system. Therefore, $J_L$ is always positive
(current flows outside) irrespective of the direction of voltage bias between the two primary dots. On the contrary, $J_R$ can become negative.

 %
\begin{figure}[tbh!]
 \centering
\includegraphics[width=1\columnwidth]  {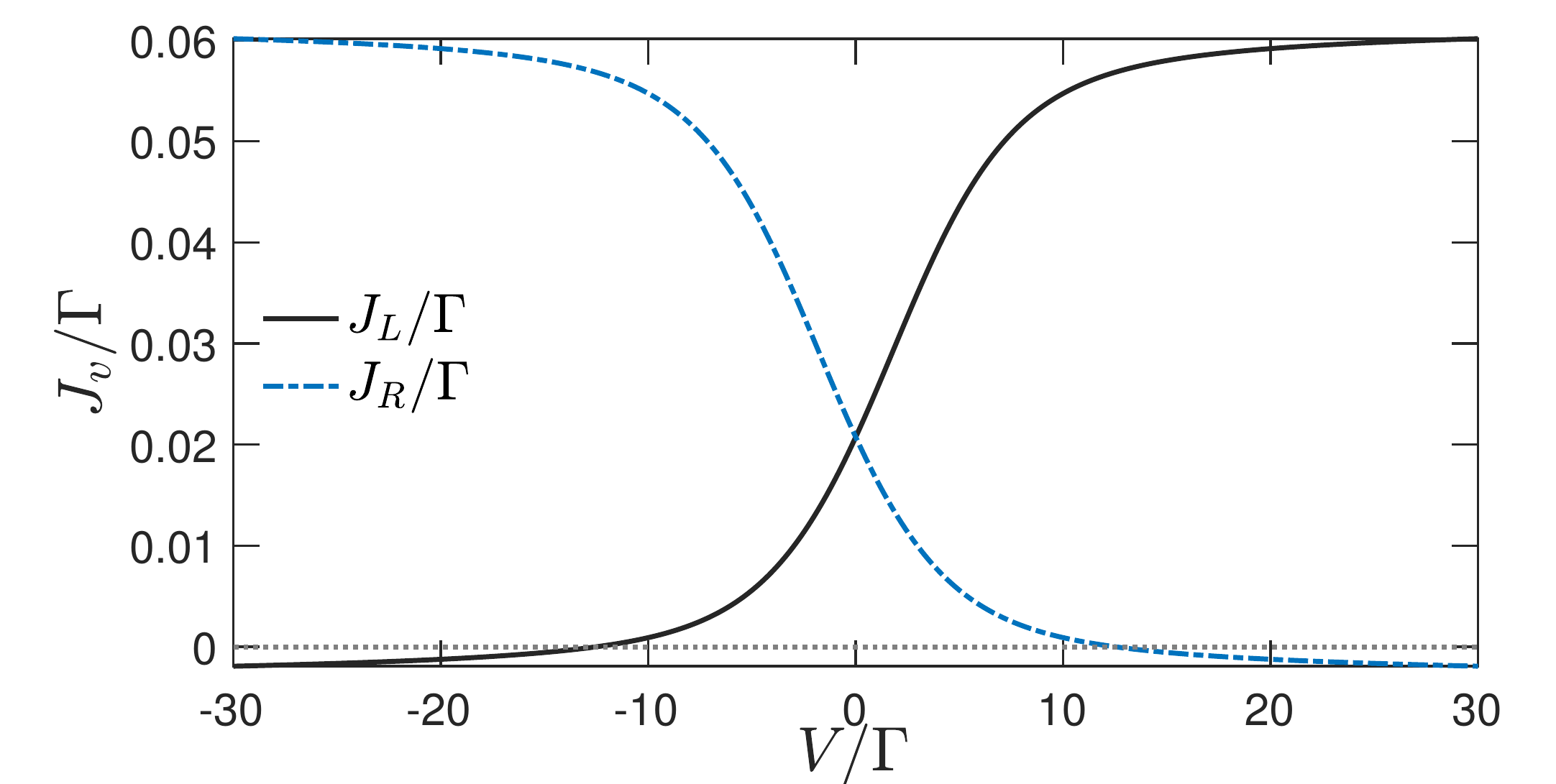} 
\caption{
Charge currents with {\it reciprocal} tunneling elements in a four-dot system.
We show the currents out of left lead ($J_L$, solid line) and right lead ($J_R$, dashed-dotted line) 
as a function of voltage bias $V$ as obtained from the Landauer-B\"uttiker formula. 
We set $\phi=\pi$, $\epsilon_d/\Gamma=1$ and $\lambda/\Gamma=2$. 
Other parameters are the same with Fig. \ref{fig:fig9}.
}
\label{fig:fig10}
\end{figure}
%
It is evident that qualitative features of Fig. \ref{fig:fig9} are similar to those of Fig. \ref{fig:fig6} for the three-dot system:
In the resonant transport regime as shown in Fig. \ref{fig:fig9} (a), electrons leaving the right lead are routed into the auxiliary dots, and electrons do not flow into the left lead in the reverse bias regime with $\mu_L<\mu_R$. While in the case of forward (positive) bias, the charge current out of the left lead can flow into the right electrode.
However, due to the presence of two auxiliary dots, only a small fraction of the electrons can be transmitted into the right lead, which is in accordance with the fact that $S_{21}$ cannot reach the maximum value 1 in this setup. In the off-resonant transport regime, depicted in Fig. \ref{fig:fig9} (b), nonreciprocal transport is largely suppressed as the negative contributions on the RHS of Eq. (\ref{eq:JR_fd}) become negligible and the first two terms on the RHS prevail. This makes $J_{L,R}$ almost symmetric, $J_L(-V)\approx J_R(V)$.

For comparison, we study the charge current with real-valued tunneling elements based on Eq.  (\ref{eq:LB}).
 For the sake of simplicity, the transmission functions are given in Appendix \ref{a:4}. 
 Results for $J_L$ and $J_R$ are shown in Fig. \ref{fig:fig10}, illustrating reciprocity.

Similar to the three-dot system,  charge currents in a fully-coherent setup are one order of magnitude smaller than those in optimal configuration shown in Fig. \ref{fig:fig9} (a). The disparity results from the fact that the latter [cf. Eqs. (\ref{eq:JL_fd}) and (\ref{eq:JR_fd})] involves only single dot transmission function $2\Gamma^2/(4\Gamma^2+(\epsilon_d-\epsilon)^2)$ and its square, which are more efficient in charge transport than those for an intrinsic four-dot system as listed in appendix \ref{a:4}. Hence, we further confirm that nonreciprocal interaction engineering not only induce unidirectional transport, but also enhance the magnitude of charge currents.


\section{Electron-phonon coupling and nonreciprocity}\label{s:4}

So far, we only considered engineering nonreciprocal interaction in noninteracting electron systems. 
For the purpose of applications, it is crucial to investigate whether the so-obtained nonreciprocal behavior persists in
 the presence of many-body interactions such as electron-phonon couplings. 
To this end, we take the three-dot system as an example. 
The coherent Hamiltonian in Eq. (\ref{eq:coh_td}) is extended to comprise collections of phonons with electron-phonon couplings,
\bea
H_M &= &H_{coh}
\nonumber\\
&+&
\sum_{n=1,2}\sum_k \left[\omega_{n,k}b_{n,k}^{\dagger}b_{n,k}
+ \gamma_{n,k}\omega_{n,k}(b_{n,k}^{\dagger}+b_{n,k})d_n^{\dagger}d_n\right]
\nonumber\\
&&+\sum_k\left[\omega_{a,k}b_{a,k}^{\dagger}b_{a,k}+
\gamma_{a,k}\omega_{a,k}(b_{a,k}^{\dagger}+b_{a,k})a^{\dagger}a \right].
\label{eq:HM}
\eea
Recall that $n=1,2$ counts primary dots and that the index $a$ corresponds to the auxiliary dot.
We assume that all dots are coupled to their local phonon environment.
Here, local phonons with frequencies $\{\omega_{l,k}\}$ ($l=1,2,a$)
are described by bosonic annihilation operators $\{b_{l,k}\}$;
electron-phonon couplings are measured by dimensionless coupling strengths, $\{\gamma_{l,k}\}$. 
The influence of each phonon bath, acting on electrons, is characterized by the spectral density function
 $I_l(\omega)=\pi\sum_k\gamma_{l,k}^2\omega_{l,k}^2\delta(\omega-\omega_{l,k})$. 
For simplicity, we assume that the three dots have the same phonon spectral density, that is, $I(\omega)=I_l(\omega)$. 
Without loss of generality, we adopt an Ohmic spectrum $I(\omega)=\pi\nu\omega e^{-\omega/\omega_c}$ with $\nu$ a dimensionless electron-phonon coupling strength and $\omega_c$ the cut-off frequency of the phonon bath.

To handle potentially strong electron-phonon couplings, we perform the small polaron transformation with the 
unitary operator $G\equiv\prod_{n=1,2}\mathcal{D}_{n}^{d_n^{\dagger}d_n}\mathcal{D}_{a}^{a^{\dagger}a}$
and displacement operators ($l=1,2,a$)
\bea
\mathcal{D}_{l}\equiv\exp\Big[\sum_k\gamma_{l,k}(b_{l,k}^{\dagger}-b_{l,k})\Big].
\label{eq:dispb}
\eea 
The transformed Hamiltonian then reads
\bea
\breve{H}_M &=& GH_MG^{\dagger}\nonumber\\
&=& \breve{\epsilon}_d\sum_{n=1,2}d_n^{\dagger}d_n-\Delta a^{\dagger}a+[g\breve{d}_1^{\dagger}\breve{d}_2+\lambda \breve{a}^{\dagger}(\breve{d}_1+\breve{d}_2)+\mathrm{H.c.}]\nonumber\\
&&+\sum_{l=1,2,a}\sum_k\omega_{l,k}b_{l,k}^{\dagger}b_{l,k}.
\eea
As can be seen, the transformation amounts to the renormalization of on-site energies, 
$\epsilon_d\to\breve{\epsilon}_{d}=\epsilon_{d}-\Delta$ with $\Delta=\int d\omega \frac{I(\omega)}{\pi\omega}$, and to the dressing of tunneling elements.
 In the above expressions, we introduced the polaron operators as 
\begin{equation}\label{eq:pf_operator}
\breve{d}_{n}~\equiv~\mathcal{D}_{n}^{\dagger}d_n,~~\breve{a}~\equiv~\mathcal{D}_{a}^{\dagger}a.
\end{equation}
In the polaron frame, the input-output relation, Eq. (\ref{eq:io_relation}), is modified as
follows \cite{Liu.20.A}
\begin{equation}\label{eq:io_pf}
\boldsymbol{F}_{out}(t)~=~\boldsymbol{F}_{in}(t)-i\sqrt{\frac{2}{\pi}}\boldsymbol{K}\cdot\breve{\boldsymbol{O}}(t).
\end{equation}
%
Recall that
$\boldsymbol{F}_{\beta}=(d_{1,\beta}, d_{2,\beta}, a_{\beta})^T$  ($\beta =in,~out$), 
$\breve{\boldsymbol{O}}=(\breve{d}_1, \breve{d}_2, \breve{a})^T$, and that $\boldsymbol{K}=\mathrm{diag}(\Gamma,\Gamma,\kappa)$. 

To obtain HLEs for electronic operators in the polaron frame, we replace the coherent Hamiltonian by $\breve{H}_M$ in Eq. (\ref{eq:eom_o}) \cite{Liu.20.A}, yielding
\bea
\label{eq:phononEOM}
\dot{d}_1 &=& -(i\breve{\epsilon}_d+\Gamma)d_1-ig\mathcal{D}_{1}\mathcal{D}_{2}^{\dagger}d_2-i\sqrt{2\pi}\mathcal{D}_{1}d_{1,in}-i\lambda \mathcal{D}_{1}\mathcal{D}_{a}^{\dagger}a,
\nonumber\\
\dot{d}_2 &=& -(i\breve{\epsilon}_d+\Gamma)d_2-ig^{\ast}\mathcal{D}_{2}\mathcal{D}_{1}^{\dagger}d_1-i\sqrt{2\pi}\mathcal{D}_{2}d_{2,in}-i\lambda \mathcal{D}_{2}\mathcal{D}_{a}^{\dagger} a,
\nonumber\\
\dot{a} &=& (i\Delta-\kappa)a-i\sqrt{2\pi} \mathcal{D}_{a}a_{in}-i\lambda(\mathcal{D}_{a}\mathcal{D}_{1}^{\dagger}d_1+\mathcal{D}_{a}\mathcal{D}_{2}^{\dagger}d_2).
\eea 
In the large damping limit, we adiabatically solve the last HLE, 
\begin{equation}\label{eq:a_hle_pf}
a~\simeq~-i\frac{\sqrt{2\pi}}{\kappa}\mathcal{D}_{a}a_{in}-i\frac{\lambda}{\kappa}(\mathcal{D}_{a}\mathcal{D}_{1}^{\dagger}d_1+\mathcal{D}_{a}\mathcal{D}_{2}^{\dagger}d_2),
\end{equation}
where we have approximated $\kappa-i\Delta\simeq\kappa$ as $\kappa\gg\Delta$. 
Inserting it into the HLEs for $d_1$, $d_2$, we find
\bea\label{eq:dd_hle_pf}
\dot{d}_1 &=& -\left(i\breve{\epsilon}_d+\Gamma+\frac{\lambda^2}{\kappa}\right) d_1-\left(\frac{\lambda^2}{\kappa}+ig\right)\mathcal{D}_{1}\mathcal{D}_{2}^{\dagger}d_2\nonumber\\
&&-i\sqrt{2\pi}\mathcal{D}_{1}d_{1,in}-\frac{\lambda}{\kappa}\sqrt{2\pi}\mathcal{D}_{1}a_{in},\nonumber\\
\dot{d}_2 &=& -\left(i\breve{\epsilon}_d+\Gamma+\frac{\lambda^2}{\kappa}\right) d_2-\left(\frac{\lambda^2}{\kappa}+ig^{\ast}\right)\mathcal{D}_{2}\mathcal{D}_{1}^{\dagger}d_1\nonumber\\
&&-i\sqrt{2\pi}\mathcal{D}_{2}d_{2,in}-\frac{\lambda}{\kappa}\sqrt{2\pi}\mathcal{D}_{2}a_{in}.
\eea
As can be seen, one can obtain perfect directionality by setting
\begin{equation}\label{eq:g_value_pf}
\frac{\lambda^2}{\kappa}+ig=0,
\end{equation}
which is precisely the directionality condition identified for a {\it noninteracting} three-dot system [cf. Eq. (\ref{eq:g_value})]. 
Hence, nonreciprocity can be achieved in the presence of local electron-phonon couplings, and the directionality condition is identical with and without
this interaction.

Next, we turn to the scattering matrix in the presence of electron-phonon couplings
and study whether the impedance matching condition, Eq. (\ref{eq:gamma_value}), holds.
As the drift matrix and the coefficient matrix in Eq. (\ref{eq:eom_general}) become time-dependent, the previous definition for the scattering matrix is no longer applicable. 
However, we now show that diagonal reflection elements $\breve{\tilde{S}}_{11}[\omega]$ and $\breve{\tilde{S}}_{22}[\omega]$ can still be identified, allowing to enforce an impedance matching condition. As a comment, we note that phonon scattering effects influence transport from dot 1 to dot 2, thus 
we cannot identify a single-frequency transmission element $\breve{\tilde{S}}_{21}[\omega]$.

In parallel to polaron operators defined in Eq. (\ref{eq:pf_operator}), we introduce the polaron-dressed input/output fields ($\beta\in in,~out$)
\begin{equation}
\breve{\boldsymbol{F}}_{\beta}=(\mathcal{D}_1d_{1,\beta}, \mathcal{D}_2d_{2,\beta}, \mathcal{D}_aa_{\beta})^T
\end{equation}
and rewrite the input-output relation Eq. (\ref{eq:io_pf}) as
\begin{equation}
\breve{\boldsymbol{F}}_{out}~=~\breve{\boldsymbol{F}}_{in}-i\sqrt{\frac{2}{\pi}}\boldsymbol{K}\cdot\boldsymbol{O}.
\label{eq:IO}
\end{equation}
We are thus seeking to understand the scattering of incoming polarons (phonon-dressed electrons) in the setup.

We can now proceed precisely as we did through Eqs. (\ref{eq:eom_general})- (\ref{eq:smatrix}).
We formally write down (\ref{eq:phononEOM}) as follows,
\begin{equation}
\label{eq:Pheom_general}
\dot{\boldsymbol{O}}(t)=\breve{\boldsymbol{M}}(t)\cdot \boldsymbol{O}(t)+\boldsymbol{C}\cdot\breve{\boldsymbol{F}}_{in}(t), 
\end{equation}
where the matrix $\breve{\boldsymbol{M}}$ generally depends on the bosonic-bath operators.
However, the equation of motion for $d_1$, 
Eq. (\ref{eq:dd_hle_pf}), does not depend on $d_2$, thus $\breve M_{1,2}=0$. 
In frequency domain (with `$\mathcal{F}$' as Fourier's transform) we write
\bea
-i\omega {\tilde d_1}[\omega] &=& -\left(i\breve{\epsilon_d } +\Gamma + \frac{\lambda^2}{\kappa} \right){\tilde d_1}[\omega]
\nonumber\\
& -&i \sqrt{2\pi} 
\breve{\tilde{d}}_{1,in}[\omega] -\frac{\lambda}{\kappa}\sqrt{2\pi}\mathcal{F}[D_1a_{in}].
\label{eq:EOMph}
\eea
%
Together with Eq. (\ref{eq:IO}), which connects the input and output fields on site `1',
we identify  the reflection element  $\breve {\tilde d}_{1,out}[\omega]=\breve{\tilde{S}}_{11}[\omega]\breve {\tilde d}_{1,in}[\omega]$ as
\begin{equation}\label{eq:s_td_pf}
\breve{\tilde{S}}_{11}[\omega]~=
~ \frac{i(\breve{\epsilon}_d-\omega)+\lambda^2/\kappa-\Gamma}{i(\breve{\epsilon}_d-\omega)+\lambda^2/\kappa+\Gamma}.
\end{equation}
Next, shifting attention to the EOM for $d_2$, we note that $M_{2,1}$ depends on bosonic operators (thus on time), leading to a convolution in the frequency domain, corresponding to phonon scattering processes. In Fourier's space we write
 \bea
-i\omega {\tilde d_2}[\omega] &=& -\left(i\breve{\epsilon_d } +\Gamma + \frac{\lambda^2}{\kappa} \right){\tilde d_2}[\omega]
 -i \sqrt{2\pi}{\breve{\tilde d}}_{2,in}[\omega]
 \nonumber\\
 &-&\frac{\lambda}{\kappa}\sqrt{2\pi} \mathcal{F} [D_2a_{in}]
- \frac{2\lambda^2}{\kappa} \mathcal{F}[D_2 D_1^{\dagger} d_1],
\eea
without explicitly writing down the last two terms.
Specifically, due to the directionality condition the last term does not depend on the input field $d_{2,in}$. Thus, together with Eq. (\ref{eq:IO}), 
one can readily extract $\breve{\tilde{S}}_{22}[\omega]$,
and further find that $\breve{\tilde{S}}_{22}[\omega]~= \breve{\tilde{S}}_{11}[\omega]$.

Eq. (\ref{eq:s_td_pf}) takes the same form as  Eq. (\ref{eq:ss}), except that the electronic energy is renormalized by electron-phonon couplings, that is, $\epsilon_d\to\breve{\epsilon}_d$.
 Hence, we recovered the impedance condition Eq. (\ref{eq:gamma_value}) with
  $\breve{\tilde{S}}_{11}[\breve{\epsilon}_d]=\breve{\tilde{S}}_{22}[\breve{\epsilon}_d]=0$.
  
\begin{figure}[tbh!]
 \centering
\includegraphics[width=1\columnwidth] {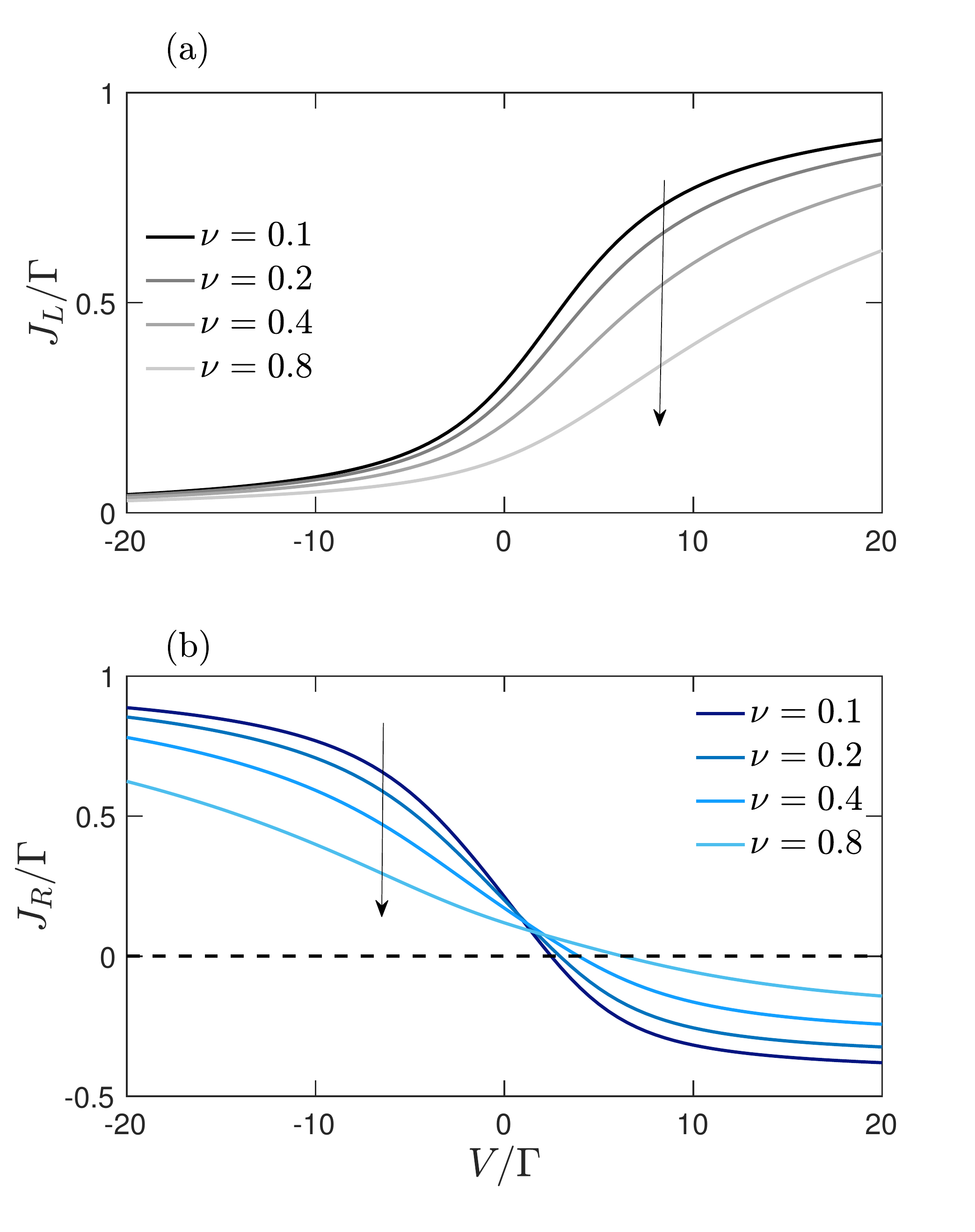} 
\caption{Charge current in the three-dot system as a function of voltage $V$, while  increasing (top to bottom) the 
dimensionless electron-phonon coupling strength $\nu$ for 
(a) $J_L$ and 
(b) $J_R$. 
The directionality condition and the impedance matching condition are fulfilled. 
We set $\omega_c/\Gamma=10$ and $\tilde{\epsilon}_d/\Gamma=1$, other parameters are the same as in Fig. \ref{fig:fig6} (a).
}
\label{fig:fig11}
\end{figure}
  Altogether, while the transport towards electrode `2' involves phonon scatterings,
importantly, diagonal elements of the scattering matrix do not depend on bosonic operators, and hence are time-independent. 
We found that  both directionality and impedance matching conditions [cf. Eqs. (\ref{eq:g_value}) and (\ref{eq:gamma_value}) ] identified in the noninteracting electron scenario survived in the presence of local electron-phonon couplings. 
 This nontrivial observation is another central result of our work.

Although with local electron-phonon couplings we retrieved the same optimal conditions for nonreciprocal interactions as
in the noninteracting case, electron-phonon interactions suerly
 affect transport of electrons (noting the off-diagonal elements of $\breve{\tilde{S}}$ involve convolutions between displacement operators and electronic ones)  and consequently the charge current, which is given by  (details can be found in Appendix \ref{a:5}),
\bea
J_L &=& 2\Gamma\int\,\frac{d\epsilon}{2\pi}\mathcal{G}(\epsilon)[n_F^L(\epsilon)-n_F^A(\epsilon)],\nonumber\\
J_R &=& 2\Gamma\int\,\frac{d\epsilon}{2\pi}\mathcal{G}(\epsilon)[n_F^R(\epsilon)-n_F^A(\epsilon)]\nonumber\\
&&-4\Gamma^2\int\,\frac{d\epsilon}{2\pi}\Big[\mathcal{G}(\epsilon)\Big]^2[n_F^L(\epsilon)-n_F^A(\epsilon)].
\eea
Here, we introduced a generalized transmission function, $\mathcal{G}(\epsilon)=\mathrm{Re}\Big[\int_{0}^{\infty}d\tau e^{-(2\Gamma+i\tilde{\epsilon}_d-i\epsilon)\tau}B(\tau)\Big]$ with $B(\tau)$ the phonon correlation function $B(\tau)=\exp\left[-\int\,d\omega\frac{I(\omega)}{\pi\omega^2}\Big(\coth(\omega/2T)(1-\cos\omega\tau)+i\sin\omega\tau\Big)\right]$. 
As can be seen, Eqs. (\ref{eq:JL}) and (\ref{eq:JR}) are recovered in the noninteracting limit where $B(\tau)=1$. 

In Fig. \ref{fig:fig11}, we show results for $J_{L, R}$ with varying the dimensionless electron-phonon coupling strength $\nu$.
Compared to Fig. \ref{fig:fig6} (a), we see that the presence of electron-phonon coupling suppresses the nonreciprocal behavior in electron transport: The magnitudes of $J_L$ and $J_R$ depict a monotonic decreasing trend as a function of $\nu$ in the whole voltage regime, in accordance with previous findings \cite{Liu.20.A}. Particularly, the contrast between $J_L(-V)$ and $J_R(V)$ reduces with increasing $\nu$.

\section{Summary}\label{s:5}

In this work, we developed an input-output scheme, which allows for exploring nonreciprocal interactions in electronic, quantum dot systems based on their quantum optomechanical analogs. In particular, Heisenberg-Langevin equations for electronic operators provide a natural definition for the scattering matrix, relating input and output fields from the metallic leads. 
With this theoretical advance,  we constructed nonreciprocity through dissipation-engineering by utilizing auxiliary damped dots and identifying
 optimal conditions for nonreciprocal interaction in a straightforward manner. 

For illustrations, we considered minimal multiterminal quantum dot models, which can serve as building blocks for larger networks. In the noninteracting 
electron scenario, we identified directionality condition as well as the so-called impedance matching condition under which optimal nonreciprocal behaviors emerged. We further showed unidirectional charge transport in the optimal regime for nonequilibrium settings, hence one can build quantum diodes by utilizing nonreciprocal interactions. We also demonstrated that electron-phonon couplings preserve optimal conditions. As we considered  metallic leads as well as solid-state quantum dot architecture, our proposal can be realized and verified by state-of-the-art experimental techniques.

In summary, the main contributions of this work for nonreciprocal electronic devices are:
(i) The GIOM formalism can be conveniently adopted to identify conditions for nonreicprocity at the level of the scattering matrix.
(ii) The electrical current, which {\it integrates} electrons in the bias window, maintains signatures of nonreciprocity in the resonant tunneling regime.
(ii) Local electron-phonon couplings preserve nonreciprocal conditions, but overall currents are gradually suppressed at strong coupling.

We note that intrinsic dissipation of auxiliary dots, which takes electron out of the system regardless of the transmission direction, is an essential ingredient for the nonreciprocal behavior reported here. However, it also reduces transmission efficiency between primary dots as can be seen from the comparison between the four-dot system and the three-dot counterpart, thereby making the use of the proposed dissipation engineering in large quantum dot networks inefficient. Further improvements are required, and we leave them to future works.

\begin{acknowledgments}
The authors acknowledge support from the Natural Sciences and Engineering Research Council (NSERC) of Canada Discovery Grant
and the Canada Research Chairs Program.
\end{acknowledgments}

\appendix
\renewcommand{\theequation}{A\arabic{equation}}
\setcounter{equation}{0}  
\section{Statistics for input fields}\label{a:1}
As the input fields depend only on initial conditions, see Eq. (\ref{eq:3}), we easily identify the following anticommutation relations
\begin{subequations}
\begin{align}
\{d_{n,in}(t),d_{n',in}^{\dagger}(t')\} &= \delta_{nn'}\Gamma\int \frac{d\epsilon}{2\pi^2}e^{-i\epsilon(t-t')},\\
\{a_{n,in}(t),a_{n',in}^{\dagger}(t')\} &= \delta_{nn'}\kappa\int \frac{d\epsilon}{2\pi^2}e^{-i\epsilon(t-t')},
\end{align}
\end{subequations}
and the following correlation functions for input fields 
\bea
&&\langle d_{n,in}(t)d_{n',in}^{\dagger}(t')\rangle~=~\delta_{nn'}\Gamma\int \frac{d\epsilon}{2\pi^2}e^{-i\epsilon(t-t')}\left[1-n_F^v(\epsilon)\right],\nonumber\\
&&\langle d_{n,in}^{\dagger}(t')d_{n',in}(t)\rangle~=~\delta_{nn'}\Gamma\int \frac{d\epsilon}{2\pi^2}e^{-i\epsilon(t-t')}n_F^v(\epsilon),\nonumber\\
&&\langle a_{m,in}(t)a_{m',in}^{\dagger}(t')\rangle~=~\delta_{mm'}\kappa\int \frac{d\epsilon}{2\pi^2}e^{-i\epsilon(t-t')}\left[1-n_F^{v'}(\epsilon)\right],\nonumber\\
&&\langle a_{m,in}^{\dagger}(t')a_{m',in}(t)\rangle~=~\delta_{mm'}\kappa\int \frac{d\epsilon}{2\pi^2}e^{-i\epsilon(t-t')}n_F^{v'}(\epsilon).
\eea
Here, $v=L (R)$ when $n=1(2)$ for the two primary dots, $v'=a$ for the three-dot system and $v'=u, d$ for the four-dot system, $n_F^v(\epsilon)=\{\exp[(\epsilon-\mu_v)/T]+1\}^{-1}$ is the Fermi-Dirac distribution with temperature $T$ and chemical potential $\mu_v$. 

\begin{widetext}
\renewcommand{\theequation}{B\arabic{equation}}
\setcounter{equation}{0}  
\section{Evaluating charge currents for the three-dot system}\label{a:2}
To get the steady state charge currents, we solve Eq. (\ref{eq:dd_hle}) under the directionality condition, Eq. (\ref{eq:g_value}) with the following stationary solutions for $d_{1,2}$ in the limit of $t_0\to-\infty$:
\bea
d_1(t) &=& -i\sqrt{2\pi}\int_{-\infty}^{t}e^{-(\Gamma+\lambda^2/\kappa+i\epsilon_d)(t-\tau)}d_{1,in}(\tau)d\tau-\frac{\lambda}{\kappa}\sqrt{2\pi}\int_{-\infty}^{t}e^{-(\Gamma+\lambda^2/\kappa+i\epsilon_d)(t-\tau)}a_{in}(\tau)d\tau,\nonumber\\
d_2(t) &=& -i\sqrt{2\pi}\int_{-\infty}^{t}e^{-(\Gamma+\lambda^2/\kappa+i\epsilon_d)(t-\tau)}d_{2,in}(\tau)d\tau-\frac{\lambda}{\kappa}\sqrt{2\pi}\int_{-\infty}^{t}e^{-(\Gamma+\lambda^2/\kappa+i\epsilon_d)(t-\tau)}a_{in}(\tau)d\tau\nonumber\\
&&-2\frac{\lambda^2}{\kappa}\int_{-\infty}^{t}e^{-(\Gamma+\lambda^2/\kappa+i\epsilon_d)(t-\tau)}d_{1}(\tau)d\tau.
\eea

We first focus on $J_L$ as $d_1$ does not depend on $d_2$. Utilizing the correlation functions for input fields listed in appendix \ref{a:1}, the involved ensemble averages in $J_L$ [see definition in Eq. (\ref{eq:J_definition})] can be evaluated as
\bea
\langle d_{1}^{\dagger}d_{1,in}\rangle &=& i\sqrt{2\pi}\Gamma\int\,\frac{d\epsilon}{2\pi^2}\frac{n_F^L(\epsilon)}{\Gamma+\lambda^2/\kappa-i\epsilon_d+i\epsilon},\nonumber\\
\langle d_{1}^{\dagger}d_{1}\rangle &=& 2\int\,\frac{d\epsilon}{2\pi}\frac{\Gamma n_F^L(\epsilon)+(\lambda^2/\kappa)n_F^A(\epsilon)}{(\Gamma+\lambda^2/\kappa)^2+(\epsilon_d-\epsilon)^2}.
\eea
Inserting them into the definition of $J_L$, we find 
\begin{equation}\label{eq:jl_b}
J_L~=~\int\,\frac{d\epsilon}{2\pi}\frac{4\Gamma\lambda^2/\kappa}{(\Gamma+\lambda^2/\kappa)^2+(\epsilon_d-\epsilon)^2}[n_F^L(\epsilon)-n_F^A(\epsilon)].
\end{equation}
Eq. (\ref{eq:JL}) of the main text is recovered by imposing the impedance matching condition $\Gamma=\lambda^2/\kappa$. 

For $J_R$, we first have,
\begin{equation}\label{eq:b2}
\langle d_{2}^{\dagger}d_{2,in}\rangle~=~i\sqrt{2\pi}\Gamma\int\,\frac{d\epsilon}{2\pi^2}\frac{n_F^R(\epsilon)}{\Gamma+\lambda^2/\kappa-i\epsilon_d+i\epsilon}.
\end{equation}
For the average occupation number, we find
\bea\label{eq:b3}
\langle d_{2}^{\dagger}d_{2}\rangle &=& 2\pi\int_{-\infty}^t\,e^{-(\Gamma+\lambda^2/\kappa-i\epsilon_d)(t-\tau)}d\tau\int_{-\infty}^t\,e^{-(\Gamma+\lambda^2/\kappa+i\epsilon_d)(t-\tau')}d\tau'\langle d_{2,in}^{\dagger}(\tau)d_{2,in}(\tau')\rangle\nonumber\\
&&+\frac{\lambda^2}{\kappa^2}2\pi\int_{-\infty}^t\,e^{-(\Gamma+\lambda^2/\kappa-i\epsilon_d)(t-\tau)}d\tau\int_{-\infty}^t\,e^{-(\Gamma+\lambda^2/\kappa+i\epsilon_d)(t-\tau')}d\tau'\langle a_{in}^{\dagger}(\tau)a_{in}(\tau')\rangle\nonumber\\
&&+4\left(\frac{\lambda^2}{\kappa}\right)^2\int_{-\infty}^t\,e^{-(\Gamma+\lambda^2/\kappa-i\epsilon_d)(t-\tau)}d\tau\int_{-\infty}^t\,e^{-(\Gamma+\lambda^2/\kappa+i\epsilon_d)(t-\tau')}d\tau'\langle d_{1}^{\dagger}(\tau)d_{1}(\tau')\rangle\nonumber\\
&&+\frac{2\lambda^3}{\kappa^2}\sqrt{2\pi}\int_{-\infty}^t\,e^{-(\Gamma+\lambda^2/\kappa-i\epsilon_d)(t-\tau)}d\tau\int_{-\infty}^t\,e^{-(\Gamma+\lambda^2/\kappa+i\epsilon_d)(t-\tau')}d\tau'\langle a_{in}^{\dagger}(\tau)d_{1}(\tau')\rangle\nonumber\\
&&+\frac{2\lambda^3}{\kappa^2}\sqrt{2\pi}\int_{-\infty}^t\,e^{-(\Gamma+\lambda^2/\kappa-i\epsilon_d)(t-\tau)}d\tau\int_{-\infty}^t\,e^{-(\Gamma+\lambda^2/\kappa+i\epsilon_d)(t-\tau')}d\tau'\langle d_1^{\dagger}(\tau)a_{in}(\tau')\rangle.
\eea
The first two terms on the RHS of Eq. (\ref{eq:b3}) can be simplified as
\begin{equation}
2\int\,\frac{d\epsilon}{2\pi}\frac{\Gamma n_F^R(\epsilon)+(\lambda^2/\kappa)n_F^A(\epsilon)}{(\Gamma+\lambda^2/\kappa)^2+(\epsilon_d-\epsilon)^2}.
\end{equation}
For the third term on the RHS, we find
\begin{equation}
8\left(\frac{\lambda^2}{\kappa}\right)^2\int\,\frac{d\epsilon}{2\pi}\frac{\Gamma n_F^L(\epsilon)+(\lambda^2/\kappa)n_F^A(\epsilon)}{[(\Gamma+\lambda^2/\kappa)^2+(\epsilon_d-\epsilon)^2]^2}
\end{equation}
by noting 
\begin{equation}\label{eq:b8}
\langle d_{1}^{\dagger}(\tau)d_{1}(\tau')\rangle~=~2\int\,\frac{d\epsilon}{2\pi}\frac{e^{-i\epsilon(\tau'-\tau)}}{(\Gamma+\lambda^2/\kappa)^2+(\epsilon_d-\epsilon)^2}[\Gamma n_F^L(\epsilon)+(\lambda^2/\kappa)n_F^A(\epsilon)].
\end{equation}
The last two terms on the RHS give 
\begin{equation}
-8\left(\frac{\lambda^2}{\kappa}\right)^2\left(\Gamma+\frac{\lambda^2}{\kappa}\right)\int\,\frac{d\epsilon}{2\pi}\frac{n_F^A(\epsilon)}{[(\Gamma+\lambda^2/\kappa)^2+(\epsilon_d-\epsilon)^2]^2}
\end{equation}
as
\bea\label{eq:b10}
\langle a_{in}^{\dagger}(\tau)d_{1}(\tau')\rangle &=& -\lambda\sqrt{2\pi}\int\,\frac{d\epsilon}{2\pi^2}n_F^A(\epsilon)\frac{e^{-i\epsilon(\tau'-\tau)}}{\Gamma+\lambda^2/\kappa+i\epsilon_d-i\epsilon},\nonumber\\
\langle d_1^{\dagger}(\tau)a_{in}(\tau')\rangle &=& -\lambda\sqrt{2\pi}\int\,\frac{d\epsilon}{2\pi^2}n_F^A(\epsilon)\frac{e^{-i\epsilon(\tau'-\tau)}}{\Gamma+\lambda^2/\kappa-i\epsilon_d+i\epsilon}.
\eea
Putting those terms together we get
\begin{equation}\label{eq:b9}
\langle d_{2}^{\dagger}d_{2}\rangle~=~\int\,\frac{d\epsilon}{2\pi}\frac{2[(\Gamma+\lambda^2/\kappa)^2+(\epsilon_d-\epsilon)^2][\Gamma n_F^R(\epsilon)+(\lambda^2/\kappa)n_F^A(\epsilon)]+8(\lambda^2/\kappa)^2\Gamma[n_F^L(\epsilon)-n_F^A(\epsilon)]}{[(\Gamma+\lambda^2/\kappa)^2+(\epsilon_d-\epsilon)^2]^2}.
\end{equation}
Inserting Eqs. (\ref{eq:b2}) and (\ref{eq:b9}) into the definition of $J_R$, we find the following expression
\bea\label{eq:jr_b}
J_R &=& \int\,\frac{d\epsilon}{2\pi}\frac{16\Gamma^2(\lambda^2/\kappa)^2}{[(\Gamma+\lambda^2/\kappa)^2+(\epsilon_d-\epsilon)^2]^2}\Big[n_F^R(\epsilon)-n_F^L(\epsilon)\Big]+\int\,\frac{d\epsilon}{2\pi}\frac{4\Gamma\lambda^2/\kappa[(\Gamma-\lambda^2/\kappa)^2+(\epsilon_d-\epsilon)^2]}{[(\Gamma+\lambda^2/\kappa)^2+(\epsilon_d-\epsilon)^2]^2}\Big[n_F^R(\epsilon)- n_F^A(\epsilon)\Big]\nonumber\\
&=& \int\,\frac{d\epsilon}{2\pi}\frac{4\Gamma\lambda^2/\kappa}{(\Gamma+\lambda^2/\kappa)^2+(\epsilon_d-\epsilon)^2}\Big[n_F^R(\epsilon)-n_F^A(\epsilon)\Big]-\int\,\frac{d\epsilon}{2\pi}\frac{16\Gamma^2(\lambda^2/\kappa)^2}{[(\Gamma+\lambda^2/\kappa)^2+(\epsilon_d-\epsilon)^2]^2}\Big[n_F^L(\epsilon)- n_F^A(\epsilon)\Big].
\eea
After the impedance matching condition, $\Gamma=\lambda^2/\kappa$, we recover  Eq. (\ref{eq:JR}) in the main text.

\renewcommand{\theequation}{C\arabic{equation}}
\setcounter{equation}{0}  
\section{Evaluating charge currents for the four-dot system}\label{a:3}

Under the directionality condition, (\ref{eq:g_value}), and the impedance matching condition (\ref{eq:gamma_value}), 
we solve Eq. (\ref{eq:dd_hle_1}) in the steady state limit,
\bea
d_1(t) &=& -i\sqrt{2\pi}\int_{-\infty}^{t}e^{-(2\Gamma+i\epsilon_d)(t-\tau)}d_{1,in}(\tau)d\tau+\frac{\lambda}{\kappa+i\delta}\sqrt{2\pi}\int_{-\infty}^{t}e^{-(2\Gamma+i\epsilon_d)(t-\tau)}a_{1,in}(\tau)d\tau,\nonumber\\
&&+\frac{\lambda}{\kappa-i\delta}\sqrt{2\pi}\int_{-\infty}^{t}e^{-(2\Gamma+i\epsilon_d)(t-\tau)}a_{2,in}(\tau)d\tau,\nonumber\\
d_2(t) &=& -i\sqrt{2\pi}\int_{-\infty}^{t}e^{-(2\Gamma+i\epsilon_d)(t-\tau)}d_{2,in}(\tau)d\tau-\frac{\lambda(\kappa-i\delta)}{(\kappa+i\delta)^2}\sqrt{2\pi}\int_{-\infty}^{t}e^{-(2\Gamma+i\epsilon_d)(t-\tau)}a_{1,in}(\tau)d\tau\nonumber\\
&&+\frac{\lambda}{\kappa-i\delta}\sqrt{2\pi}\int_{-\infty}^{t}e^{-(2\Gamma+i\epsilon_d)(t-\tau)}a_{2,in}(\tau)d\tau-2\Gamma\frac{i\delta}{\kappa+i\delta}\int_{-\infty}^{t}e^{-(2\Gamma+i\epsilon_d)(t-\tau)}d_{1}(\tau)d\tau.
\eea

We first focus on $J_L$ as $d_1$ does not depend on $d_2$. By noting the correlation functions for input fields listed in appendix \ref{a:1}, the involved ensemble averages in $J_L$  [see definition in Eq. (\ref{eq:J_definition})] can be evaluated as
\bea
\langle d_{1}^{\dagger}d_{1,in}\rangle &=& i\sqrt{2\pi}\Gamma\int\,\frac{d\epsilon}{2\pi^2}\frac{n_F^L(\epsilon)}{2\Gamma-i\epsilon_d+i\epsilon},\nonumber\\
\langle d_{1}^{\dagger}d_{1}\rangle &=& 2\int\,\frac{d\epsilon}{2\pi}\frac{\Gamma n_F^L(\epsilon)+(\lambda^2\kappa/(\kappa^2+\delta^2))[n_F^u(\epsilon)+n_F^d(\epsilon)]}{4\Gamma^2+(\epsilon_d-\epsilon)^2}.
\eea
Inserting them into the definition of $J_L$ we recover Eq. (\ref{eq:JL_fd}) in the main text by noting $\Gamma=2\lambda^2\kappa/(\kappa^2+\delta^2)$. 

For $J_R$, we first have
\begin{equation}
\langle d_{2}^{\dagger}d_{2,in}\rangle~=~i\sqrt{2\pi}\Gamma\int\,\frac{d\epsilon}{2\pi^2}\frac{n_F^R(\epsilon)}{2\Gamma-i\epsilon_d+i\epsilon}.
\end{equation}
For the average occupation number, we find
\bea\label{eq:c3}
\langle d_{2}^{\dagger}d_{2}\rangle &=& 2\pi\int_{-\infty}^t\,e^{-(2\Gamma-i\epsilon_d)(t-\tau)}d\tau\int_{-\infty}^t\,e^{-(2\Gamma+i\epsilon_d)(t-\tau')}d\tau'\langle d_{2,in}^{\dagger}(\tau)d_{2,in}(\tau')\rangle\nonumber\\
&&+\frac{\lambda^2}{\kappa^2+\delta^2}2\pi\int_{-\infty}^t\,e^{-(2\Gamma-i\epsilon_d)(t-\tau)}d\tau\int_{-\infty}^t\,e^{-(2\Gamma+i\epsilon_d)(t-\tau')}d\tau'\langle a_{1,in}^{\dagger}(\tau)a_{1,in}(\tau')\rangle\nonumber\\
&&+\frac{\lambda^2}{\kappa^2+\delta^2}2\pi\int_{-\infty}^t\,e^{-(2\Gamma-i\epsilon_d)(t-\tau)}d\tau\int_{-\infty}^t\,e^{-(2\Gamma+i\epsilon_d)(t-\tau')}d\tau'\langle a_{2,in}^{\dagger}(\tau)a_{2,in}(\tau')\rangle\nonumber\\
&&+4\Gamma^2\frac{\delta^2}{\kappa^2+\delta^2}\int_{-\infty}^t\,e^{-(2\Gamma-i\epsilon_d)(t-\tau)}d\tau\int_{-\infty}^t\,e^{-(2\Gamma+i\epsilon_d)(t-\tau')}d\tau'\langle d_{1}^{\dagger}(\tau)d_{1}(\tau')\rangle\nonumber\\
&&+\frac{2\Gamma\lambda i\delta}{(\kappa-i\delta)^2}\sqrt{2\pi}\int_{-\infty}^t\,e^{-(2\Gamma-i\epsilon_d)(t-\tau)}d\tau\int_{-\infty}^t\,e^{-(2\Gamma+i\epsilon_d)(t-\tau')}d\tau'\langle a_{1,in}^{\dagger}(\tau)d_{1}(\tau')\rangle\nonumber\\
&&-\frac{2\Gamma\lambda i\delta}{(\kappa+i\delta)^2}\sqrt{2\pi}\int_{-\infty}^t\,e^{-(2\Gamma-i\epsilon_d)(t-\tau)}d\tau\int_{-\infty}^t\,e^{-(2\Gamma+i\epsilon_d)(t-\tau')}d\tau'\langle d_1^{\dagger}(\tau)a_{1,in}(\tau')\rangle\nonumber\\
&&-\frac{2\Gamma\lambda i\delta}{(\kappa+i\delta)^2}\sqrt{2\pi}\int_{-\infty}^t\,e^{-(2\Gamma-i\epsilon_d)(t-\tau)}d\tau\int_{-\infty}^t\,e^{-(2\Gamma+i\epsilon_d)(t-\tau')}d\tau'\langle a_{2,in}^{\dagger}(\tau)d_{1}(\tau')\rangle\nonumber\\
&&+\frac{2\Gamma\lambda i\delta}{(\kappa-i\delta)^2}\sqrt{2\pi}\int_{-\infty}^t\,e^{-(2\Gamma-i\epsilon_d)(t-\tau)}d\tau\int_{-\infty}^t\,e^{-(2\Gamma+i\epsilon_d)(t-\tau')}d\tau'\langle d_1^{\dagger}(\tau)a_{2,in}(\tau')\rangle.
\eea
The first three terms on the RHS of Eq. (\ref{eq:c3}) can be simplified as
\begin{equation}
2\int\,\frac{d\epsilon}{2\pi}\frac{\Gamma n_F^R(\epsilon)+(\lambda^2\kappa/(\kappa^2+\delta^2))[n_F^u(\epsilon)+n_F^d(\epsilon)]}{4\Gamma^2+(\epsilon_d-\epsilon)^2}.
\end{equation}
For the fourth term on the RHS of Eq. (\ref{eq:c3}), we find
\begin{equation}
8\Gamma^2\frac{\delta^2}{\kappa^2+\delta^2}\int\,\frac{d\epsilon}{2\pi}\frac{\Gamma n_F^L(\epsilon)+[\lambda^2\kappa/(\kappa^2+\delta^2)][n_F^u(\epsilon)+n_F^d(\epsilon)]}{[4\Gamma^2+(\epsilon_d-\epsilon)^2]^2}
\end{equation}
by noting 
\begin{equation}
\langle d_{1}^{\dagger}(\tau)d_{1}(\tau')\rangle~=~2\int\,\frac{d\epsilon}{2\pi}\frac{e^{-i\epsilon(\tau'-\tau)}}{4\Gamma^2+(\epsilon_d-\epsilon)^2}\Big\{\Gamma n_F^L(\epsilon)+[\lambda^2\kappa/(\kappa^2+\delta^2)][n_F^u(\epsilon)+n_F^d(\epsilon)]\Big\}.
\end{equation}
The last four terms on the RHS of Eq. (\ref{eq:c3}) give 
\begin{equation}
\frac{8\Gamma\lambda^2\kappa\delta}{(\kappa^2+\delta^2)^2}\int\,\frac{d\epsilon}{2\pi}\frac{\kappa(\epsilon_d-\epsilon)[n_F^u(\epsilon)-n_F^d(\epsilon)]-2\Gamma\delta[n_F^u(\epsilon)+n_F^d(\epsilon)]}{[4\Gamma^2+(\epsilon_d-\epsilon)^2]^2}
\end{equation}
by using the following correlation functions
\bea
\langle a_{1,in}^{\dagger}(\tau)d_{1}(\tau')\rangle &=& \frac{\lambda\kappa}{\kappa+i\delta}\sqrt{2\pi}\int\,\frac{d\epsilon}{2\pi^2}n_F^u(\epsilon)\frac{e^{-i\epsilon(\tau'-\tau)}}{2\Gamma+i\epsilon_d-i\epsilon},\nonumber\\
\langle d_1^{\dagger}(\tau)a_{1,in}(\tau')\rangle &=& \frac{\lambda\kappa}{\kappa-i\delta}\sqrt{2\pi}\int\,\frac{d\epsilon}{2\pi^2}n_F^u(\epsilon)\frac{e^{-i\epsilon(\tau'-\tau)}}{2\Gamma-i\epsilon_d+i\epsilon},\nonumber\\
\langle a_{2,in}^{\dagger}(\tau)d_{1}(\tau')\rangle &=& \frac{\lambda\kappa}{\kappa-i\delta}\sqrt{2\pi}\int\,\frac{d\epsilon}{2\pi^2}n_F^d(\epsilon)\frac{e^{-i\epsilon(\tau'-\tau)}}{2\Gamma+i\epsilon_d-i\epsilon},\nonumber\\
\langle d_1^{\dagger}(\tau)a_{2,in}(\tau')\rangle &=& \frac{\lambda\kappa}{\kappa+i\delta}\sqrt{2\pi}\int\,\frac{d\epsilon}{2\pi^2}n_F^d(\epsilon)\frac{e^{-i\epsilon(\tau'-\tau)}}{2\Gamma-i\epsilon_d+i\epsilon}.
\eea
Inserting $\langle d_{2}^{\dagger}d_{2,in}\rangle$ and $\langle d_{2}^{\dagger}d_{2}\rangle$ into the definition, we recover Eq. (\ref{eq:JR_fd}) in the main text under the impedance matching condition $\Gamma=2\lambda^2\kappa/(\kappa^2+\delta^2)$.

\renewcommand{\theequation}{D\arabic{equation}}
\setcounter{equation}{0}  
\section{Transmission functions for the four-dot system}\label{a:4}

We organize primary and auxiliary dots in the following order, $(1,2, u, d)$.  
The dot-lead coupling matrices read $\boldsymbol{\Gamma}_L=\mathrm{diag}(\Gamma,0,0,0)$, $\boldsymbol{\Gamma}_R=\mathrm{diag}(0,\Gamma,0,0)$, $\boldsymbol{\Gamma}_u=\mathrm{diag}(0,0,\kappa,0)$ and $\boldsymbol{\Gamma}_d=\mathrm{diag}(0,0,0,\kappa)$. The matrix form for the coherent Hamiltonian takes the following form
\begin{equation}
\boldsymbol{H}_{coh}(\phi)~=~\left(
\begin{array}{cccc}
\epsilon_d & 0 & \lambda & \lambda\\
0 & \epsilon_d & \lambda e^{i\phi} & \lambda\\
\lambda & \lambda e^{-i\phi} & \delta & 0\\
\lambda & \lambda & 0 & -\delta
\end{array}
\right)
\end{equation}
A direct calculation leads to:
\bea\label{eq:tt_fd}
\mathcal{T}_{LR}(\epsilon,\phi) &=& \mathcal{T}_{RL}(\epsilon,-\phi)~=~\frac{\Gamma^2\lambda^4|(e^{-i\phi}+1)(\epsilon+i\kappa)+(e^{-i\phi}-1)\delta|^2}{|(\epsilon-\epsilon_d-i\Gamma)^2[(\epsilon-i\kappa)^2-\delta^2]-4\lambda^2(\epsilon-\epsilon_d-i\Gamma)(\epsilon-i\kappa)-2\lambda^4(\cos\phi-1)|^2},\nonumber\\
\mathcal{T}_{Lu}(\epsilon,\phi) &=& \mathcal{T}_{uL}(\epsilon,-\phi)~=~\frac{\Gamma\kappa\lambda^2|\lambda^2(e^{-i\phi}-1)+(\epsilon+\delta-i\kappa)(\epsilon-\epsilon_d-i\Gamma)|^2}{|(\epsilon-\epsilon_d-i\Gamma)^2[(\epsilon-i\kappa)^2-\delta^2]-4\lambda^2(\epsilon-\epsilon_d-i\Gamma)(\epsilon-i\kappa)-2\lambda^4(\cos\phi-1)|^2},\nonumber\\
\mathcal{T}_{Ld}(\epsilon,\phi) &=& \mathcal{T}_{dL}(\epsilon,-\phi)~=~\frac{\Gamma\kappa\lambda^2|\lambda^2(e^{-i\phi}-1)+(\epsilon-\delta+i\kappa)(\epsilon-\epsilon_d+i\Gamma)|^2}{|(\epsilon-\epsilon_d-i\Gamma)^2[(\epsilon-i\kappa)^2-\delta^2]-4\lambda^2(\epsilon-\epsilon_d-i\Gamma)(\epsilon-i\kappa)-2\lambda^4(\cos\phi-1)|^2},\nonumber\\
\mathcal{T}_{Ru}(\epsilon,\phi) &=& \mathcal{T}_{uR}(\epsilon,-\phi)~=~\frac{\Gamma\kappa\lambda^2|\lambda^2(1-e^{-i\phi})+e^{-i\phi}(\epsilon+\delta-i\kappa)(\epsilon-\epsilon_d-i\Gamma)|^2}{|(\epsilon-\epsilon_d-i\Gamma)^2[(\epsilon-i\kappa)^2-\delta^2]-4\lambda^2(\epsilon-\epsilon_d-i\Gamma)(\epsilon-i\kappa)-2\lambda^4(\cos\phi-1)|^2},\nonumber\\
\mathcal{T}_{Rd}(\epsilon,\phi) &=& \mathcal{T}_{dR}(\epsilon,-\phi)~=~\frac{\Gamma\kappa\lambda^2|\lambda^2(e^{-i\phi}-1)+(\epsilon-\delta-i\kappa)(\epsilon-\epsilon_d-i\Gamma)|^2}{|(\epsilon-\epsilon_d-i\Gamma)^2[(\epsilon-i\kappa)^2-\delta^2]-4\lambda^2(\epsilon-\epsilon_d-i\Gamma)(\epsilon-i\kappa)-2\lambda^4(\cos\phi-1)|^2}.
\eea

\renewcommand{\theequation}{E\arabic{equation}}
\setcounter{equation}{0}  
\section{Evaluating charge currents for the three-dot system in the presence of electron-phonon couplings}\label{a:5}
In the polaron frame, the charge current definitions are modified as \cite{Liu.20.A}
\bea\label{eq:J_definition_pf}
J_L &=& 2\left(\sqrt{2\pi}\mathrm{Im}\langle \breve{d}_{1}^{\dagger}d_{1,in}\rangle-\Gamma\langle d_{1}^{\dagger}d_{1}\rangle\right),\nonumber\\
J_R &=& 2\left(\sqrt{2\pi}\mathrm{Im}\langle \breve{d}_{2}^{\dagger}d_{2,in}\rangle-\Gamma\langle d_{2}^{\dagger}d_{2}\rangle\right).
\eea
To get the steady state charge currents, we solve Eq. (\ref{eq:dd_hle_pf}) under the directionality condition 
(\ref{eq:g_value})
and the impedance matching condition (\ref{eq:gamma_value})
 in the limit of $t_0\to-\infty$:
\bea
d_1(t) &=& -i\sqrt{2\pi}\int_{-\infty}^{t}e^{-(2\Gamma+i\tilde{\epsilon}_d)(t-\tau)}\mathcal{D}_1(\tau)d_{1,in}(\tau)d\tau-\frac{\lambda}{\kappa}\sqrt{2\pi}\int_{-\infty}^{t}e^{-(2\Gamma+i\tilde{\epsilon}_d)(t-\tau)}\mathcal{D}_1(\tau)a_{in}(\tau)d\tau,\nonumber\\
d_2(t) &=& -i\sqrt{2\pi}\int_{-\infty}^{t}e^{-(2\Gamma+i\tilde{\epsilon}_d)(t-\tau)}\mathcal{D}_2(\tau)d_{2,in}(\tau)d\tau-\frac{\lambda}{\kappa}\sqrt{2\pi}\int_{-\infty}^{t}e^{-(2\Gamma+i\tilde{\epsilon}_d)(t-\tau)}\mathcal{D}_2(\tau)a_{in}(\tau)d\tau\nonumber\\
&&-2\Gamma\int_{-\infty}^{t}e^{-(2\Gamma+i\tilde{\epsilon}_d)(t-\tau)}\mathcal{D}_2(\tau)\mathcal{D}_1^{\dagger}(\tau)d_{1}(\tau)d\tau.
\eea
Similar to appendix \ref{a:2}, we first evaluate ensemble averages involved in $J_L$,
\bea
\langle \breve{d}_{1}^{\dagger}d_{1,in}\rangle &=& i\sqrt{2\pi}\int\,d\tau e^{-(2\Gamma-i\tilde{\epsilon}_d)(t-\tau)}\langle d_{1,in}^{\dagger}(\tau)\mathcal{D}_1^{\dagger}(\tau)D_1(t)d_{1,in}(t)\rangle\nonumber\\
&\approx& i\sqrt{2\pi}\Gamma\int\,d\epsilon\frac{n_F^L(\epsilon)}{2\pi^2}\int_0^{\infty}e^{-(2\Gamma-i\tilde{\epsilon}_d+i\epsilon)\tau}B^{\ast}(\tau)d\tau.
\eea
In the second line, we have decoupled electron and phonon correlations, a common approximation in the polaron frame \cite{Liu.20.A,Galperin.06.PRBa,Maier.11.PRB,Souto.15.PRB}, which relies on this coupling to be moderate,
\begin{equation}
\langle d_{1,in}^{\dagger}(\tau)\mathcal{D}_1^{\dagger}(\tau)D_1(t)d_{1,in}(t)\rangle~\approx~\langle d_{1,in}^{\dagger}(\tau)d_{1,in}(t)\rangle\langle\mathcal{D}_1^{\dagger}(\tau)D_1(t)\rangle.
\end{equation}
Here, we introduced the phonon correlation function $B(t-\tau)=\langle \mathcal{D}^{\dagger}(t)D(\tau)\rangle$; note that 
we neglect the dot subscript since the phonon environments are assumed to have identical spectral density functions, for the sake of simplicity. 
By assuming an initial thermal equilibrium state for the phonon environments, we recover the standard form
\begin{equation}
B(\tau)~=~\exp\left[-\int\,d\omega\frac{I(\omega)}{\pi\omega^2}\Big(\coth(\omega/2T)(1-\cos\omega\tau)+i\sin\omega\tau\Big)\right].
\end{equation}
Similarly, we find
\bea
\langle d_{1}^{\dagger}d_{1}\rangle &=& 2\Gamma\int\,d\epsilon\frac{n_F^L(\epsilon)+n_F^A(\epsilon)}{2\pi}\int_{-\infty}^{t}d\tau\int_{-\infty}^{t}d\tau'e^{i(\epsilon-\tilde{\epsilon}_d)(\tau-\tau')}e^{-2\Gamma(2t-\tau-\tau')}B(\tau-\tau')\nonumber\\
&=& 4\Gamma\int\,d\epsilon\frac{n_F^L(\epsilon)+n_F^A(\epsilon)}{2\pi}\mathrm{Re}\Big[\int_{-\infty}^{t}d\tau\int_{-\infty}^{\tau}d\tau'e^{i(\epsilon-\tilde{\epsilon}_d)(\tau-\tau')}e^{-2\Gamma(2t-\tau-\tau')}B(\tau-\tau')\Big]\nonumber\\
&=& \int\,d\epsilon\frac{n_F^L(\epsilon)+n_F^A(\epsilon)}{2\pi}\mathrm{Re}\Big[\int_{0}^{\infty}d\tau e^{-(2\Gamma+i\tilde{\epsilon}_d-i\epsilon)\tau}B(\tau)\Big].
\eea
Here, ``Re" refers to the real part. Inserting them into the definition of $J_L$, we find
\begin{equation}\label{eq:e7}
J_L~=~2\Gamma\int\,\frac{d\epsilon}{2\pi}\mathrm{Re}\Big[\int_{0}^{\infty}d\tau e^{-(2\Gamma+i\tilde{\epsilon}_d-i\epsilon)\tau}B(\tau)\Big][n_F^L(\epsilon)-n_F^A(\epsilon)].
\end{equation}
We note that Eq. (\ref{eq:JL}) in the main text is recovered when $B(\tau)=1$, namely, in the coherent limit of noninteracting electrons.

As for ensemble averages involved in $J_R$, we follow similar procedures and find
\begin{equation}
\langle d_{2}^{\dagger}d_{2,in}\rangle~=~i\sqrt{2\pi}\Gamma\int\,d\epsilon\frac{n_F^R(\epsilon)}{2\pi^2}\int_0^{\infty}e^{-(2\Gamma-i\tilde{\epsilon}_d+i\epsilon)\tau}B^{\ast}(\tau)d\tau,
\end{equation}
and
\bea\label{eq:e9}
\langle d_{2}^{\dagger}d_{2}\rangle &=& \int\,d\epsilon\frac{n_F^R(\epsilon)+n_F^A(\epsilon)}{2\pi}\mathrm{Re}\Big[\int_{0}^{\infty}d\tau e^{-(2\Gamma+i\tilde{\epsilon}_d-i\epsilon)\tau}B(\tau)\Big]\nonumber\\
&&+2\Gamma\int\,\frac{d\epsilon}{2\pi}[n_F^L(\epsilon)-n_F^A(\epsilon)]\Bigg(\mathrm{Re}\Big[\int_{0}^{\infty}d\tau e^{-(2\Gamma+i\tilde{\epsilon}_d-i\epsilon)\tau}B(\tau)\Big]\Bigg)^2.
\eea
Consequently, we get the following expression for $J_R$:
\bea\label{eq:e10}
J_R &=& 2\Gamma\int\,\frac{d\epsilon}{2\pi}\mathrm{Re}\Big[\int_{0}^{\infty}d\tau e^{-(2\Gamma+i\tilde{\epsilon}_d-i\epsilon)\tau}B(\tau)\Big][n_F^R(\epsilon)-n_F^A(\epsilon)]\nonumber\\
&&-4\Gamma^2\int\,\frac{d\epsilon}{2\pi}\Bigg(\mathrm{Re}\Big[\int_{0}^{\infty}d\tau e^{-(2\Gamma+i\tilde{\epsilon}_d-i\epsilon)\tau}B(\tau)\Big]\Bigg)^2[n_F^L(\epsilon)-n_F^A(\epsilon)].
\eea
The noninteracting expression, Eq. (\ref{eq:JR}) in the main text is recovered when $B(\tau)=1$.

\end{widetext}

%

\end{document}